\documentclass[sigconf]{acmart}

\def\BibTeX{{\rm B\kern-.05em{\sc i\kern-.025em b}\kern-.08emT\kern-.1667em\lower.7ex\hbox{E}\kern-.125emX}}

\usepackage{tikz}
\usepackage{amsmath}

\usepackage{graphicx}
\usepackage{subfigure}
\usepackage{multirow}
\usepackage{multicol}
\usepackage{booktabs}
\usepackage{threeparttable}
\usepackage{xcolor}
\usepackage[linesnumbered,ruled,vlined]{algorithm2e}

\usepackage{balance}
\usepackage{float}
\usepackage{adjustbox}
\usepackage{url}
\usepackage{color}
\usepackage{xspace}
\usepackage[T1]{fontenc}

\newcommand{\eat}[1]{}
\newcommand{\hide}[1]{}

\usepackage{hyperref} 
\hypersetup{
    citebordercolor={0 1 0},
    citecolor=blue,
    colorlinks=false,
    filebordercolor={0 .5 .5},
    filecolor=green,
    linkbordercolor={1 0 0},
    linkcolor=blue,
    menubordercolor={1 0 0},
    pageanchor=true,
    pagebackref=true,
    pdfborder={0 0 1},
    pdfpagelabels=true,
    pdftex,
    plainpages=false,
    urlbordercolor={0 1 1},
    urlcolor=blue,
}

\newcommand{\eg}{e.g., }
\newcommand{\ie}{i.e., }
\newcommand{\etal}{et al. }

\newcommand{\yy}[1]{{\color{red} YY: #1}}
\newcommand{\wt}[1]{{\color{red} WT: #1}}

\newcommand{\tool} {{\textsc{FTrojan}}\xspace}
\newcommand{\abs} {{\textsc{ABS}}\xspace}
\newcommand{\strip} {{\textsc{STRIP}}\xspace}
\newcommand{\nc} {{\textsc{Neural Cleanse}}\xspace}
\newcommand{\feb} {{\textsc{Februus}}\xspace}
\newcommand{\nad} {{\textsc{NAD}}\xspace}
\newcommand{\badnet} {{\textsc{BadNet}}\xspace}
\newcommand{\sig} {{\textsc{SIG}}\xspace}
\newcommand{\iab} {{\textsc{IAB}}\xspace}
\newcommand{\refool} {{\textsc{Refool}}\xspace}

\begin{document}

\title{Backdoor Attack through Frequency Domain}

\author{Tong Wang$^1$, Yuan Yao$^1$, Feng Xu$^1$, Shengwei An$^2$, Hanghang Tong$^3$,  Ting Wang$^4$}
\affiliation{\institution{$^1$ State Key Laboratory for Novel Software Technology, Nanjing University, China}}
\affiliation{\institution{$^2$ Purdue University, USA}}
\affiliation{\institution{$^3$ University of Illinois Urbana-Champaign, USA}}
\affiliation{\institution{$^4$ Pennsylvania State University, USA}}
\email{mg20330065@smail.nju.edu.cn, {xf, y.yao}@nju.edu.cn, an93@purdue.edu, htong@illinois.edu, inbox.ting@gmail.com}

\hide{
\author {
    Tong Wang, Yuan Yao, Feng Xu \textsuperscript{\rm 1}
    Shengwei An, \textsuperscript{\rm 2}
    Ting Wang, \textsuperscript{\rm 3}
}
\affiliations {
    \textsuperscript{\rm 1} State Key Laboratory for Novel Software Technology, Nanjing University, China\\
    \textsuperscript{\rm 3} Purdue University\\
    \textsuperscript{\rm 4} Pennsylvania State University\\
    mg20330065@smail.nju.edu.cn, {xf, y.yao}@nju.edu.cn, an93@purdue.edu, inbox.ting@gmail.com
}
}

\begin{abstract}
Backdoor attacks have been shown to be a serious threat against deep learning systems such as biometric authentication and autonomous driving. 
An effective backdoor attack could enforce the model misbehave under certain predefined conditions, i.e., {\em triggers}, but behave normally otherwise. 
However, the triggers of existing attacks are directly injected in the pixel space, which tend to be detectable by existing defenses and visually identifiable at both training and inference stages. 
In this paper, we propose a new backdoor attack \tool through trojaning the frequency domain. The key intuition is that triggering perturbations in the frequency domain correspond to small pixel-wise perturbations dispersed across the entire image, breaking the underlying assumptions of existing defenses and making the poisoning images visually indistinguishable from clean ones.
We evaluate \tool in several datasets and tasks showing that it achieves a high attack success rate without significantly degrading the prediction accuracy on benign inputs. Moreover, the poisoning images are nearly invisible and retain high perceptual quality.  
We also evaluate \tool against state-of-the-art defenses as well as several adaptive defenses that are designed on the frequency domain. The results show that \tool can robustly elude or significantly degenerate the performance of these defenses.

\end{abstract}

\maketitle


\section{Introduction}
Convolutional neural networks (CNNs) have attracted tremendous attention and have been widely used in many real applications including object classification~\cite{krizhevsky2012imagenet,he2015delving}, face recognition~\cite{parkhi2015deep,schroff2015facenet}, real-time object detection~\cite{ren2015faster,redmon2016you}, etc. 
To build their own CNNs, users usually need to collect a large-scale dataset from the open Internet, or even outsource the entire training process due to the lack of computing resources. This makes CNNs exploitable to backdoor/trojan attacks~\cite{gu2017badnets,liu2018trojaning}.
Specifically, a typical backdoor attack poisons a small subset of training data with a {\em trigger}, and enforces the backdoored model misbehave (\eg misclassify the test input to a target label) when the trigger is present but behave normally otherwise at inference time.
Such attacks can cause serious damages such as deceiving biometric authentication that is based on face recognition or misleading autonomous cars that rely on camera inputs.

An ideal backdoor attack should satisfy the three desiderata of {\em efficacy}, {\em specificity}, and {\em fidelity} from the adversary's perspective~\cite{pang2020tale}. Here, efficacy means that the target CNN model can be successfully misled by the triggers, specificity means that the trained model should perform normally on the benign inputs, and fidelity means the poisoning images should retain the perceptual similarity to the original clean images. The latter two aspects are related to the {\em stealthiness} of a backdoor attack. That is, if either the trigger is clearly visible or the backdoored model performs relatively poor on the benign inputs, users may easily detect such an anomaly.

While various existing backdoor attacks perform relatively well on the efficacy and specificity aspects, they tend to fall short in terms of satisfying the fidelity requirement, \ie the triggers are visually identifiable. The fundamental reason is that existing attacks directly inject or search for triggers in the spatial domain (\ie pixel space) of an image. In this domain, it is a dilemma to find triggers that are simultaneously recognizable by CNNs and invisible to humans (please refer to Figure~\ref{fig:motivation_backdoor_attacks} and Section~\ref{sec:motivate} for details). Additionally, existing triggers in the spatial domain are usually small in size and concentrated in energy, making them detectable by existing defenses~\cite{wang2019neural,liu2019abs,doan2020februus}.

In this paper, we propose a new backdoor attack \tool through the frequency domain of images. Our key insights are two-fold. First, adding small perturbations in the mid- and high-frequency components of an image will not significantly reduce its fidelity~\cite{sonka2014image,yamaguchi2018high}.\footnote{Intuitively, while low-frequency components correspond to large and flat areas in the image, mid- and high-frequency components describe the sharply changing areas (\eg edges, contours, or noises) and thus their small perturbations are nearly negligible to human perception.} 
Second, recent research has provided evidence that images' frequency-domain features are recognizable and learnable by CNNs~\cite{yin2019fourier,xu2019training,xu2019frequency,wang2020high}. In other words, although CNNs take the spatial-domain pixels as input, they can also easily learn and remember the injected triggers in the frequency domain. 

Armed with the above insights, it is still challenging to make our attack more invisible and more robust against existing defenses. For such purposes, we first transform the images from RGB channels to YUV channels as UV channels correspond to chrominance components that are less sensitive to the human visual system (HVS). Next, we divide an image into a set of disjoint blocks and inject the trigger at both mid- or high-frequency components of the UV channels in each block.
Through the above design, we can not only maintain the high fidelity of poisoning images, but also disperse the trigger throughout the entire image breaking the underlying assumptions of many existing defenses.
We also extend the attack to the clean-label setting~\cite{turner2018clean}, making the attack evasive at both training and inference stages.



%

We evaluate our attack in several datasets and tasks including traffic sign recognition, objection classification, and face recognition. 
The results show that the proposed attack \tool achieves 98.78\% attack success rate on average without significantly degrading the classification accuracy on benign inputs (0.56\% accuracy decrease on average). Moreover, we compare the fidelity aspect with several existing backdoor attacks and show that the poisoning images by \tool are nearly invisible and retain higher perceptual quality. 
We also evaluate the proposed attack against five state-of-the-art backdoor defensing systems including \nc~\cite{wang2019neural}, \abs~\cite{liu2019abs}, \strip~\cite{gao2019strip}, \feb~\cite{doan2020februus}, and \nad~\cite{li2021neural}. We also design several adaptive defenses based on anomaly detection and signal smoothing in the frequency domain. The results show that \tool can robustly bypass or significantly degenerate the performance of these defenses. 

\hide{
To be specific, the poisoning process of \tool works as follows. For a given image that \tool aims to poison, it first converts the images from RGB channels to YUV channels and injects triggers in the UV channels. The intuition is that the most sensitive information to the HVS is contained in the Y channel, and thus injecting triggers at UV channels would be less visible.
Next, \tool divides the input image into a set of disjoint blocks and performs discrete cosine transform (DCT)~\cite{ahmed1974discrete} on each block to transform the image from the spatial domain to the frequency domain.
\tool then generates and injects the trigger in the frequency domain. In particular, \tool generates triggers with moderate magnitude and injects the triggers at mid and high frequency bands over the entire image. 
By doing so, \tool could satisfy the three requirements of efficacy, specificity, and fidelity, while being able to bypass or significantly reduce the effectiveness of existing defenses against backdoor attacks.
Finally, \tool transforms the image back to the spatial domain and the RGB channels.
}

The main contributions of this paper include:
\begin{itemize}
    \item We are the first to propose backdoor attacks through trojaning the frequency domain. It opens the door for various future backdoor attacks and defenses. 
    \item We explore a large design space of backdoor attacks in frequency domain and report our findings.
    \item We empirically show the superiority of the proposed attack in terms of efficacy, specificity, and fidelity aspects, as well as the robustness against existing defenses.
\end{itemize}


\section{Motivation}\label{sec:motivate}

\hide{
\begin{figure}[t]
\centering
  \includegraphics[width=0.65\linewidth]{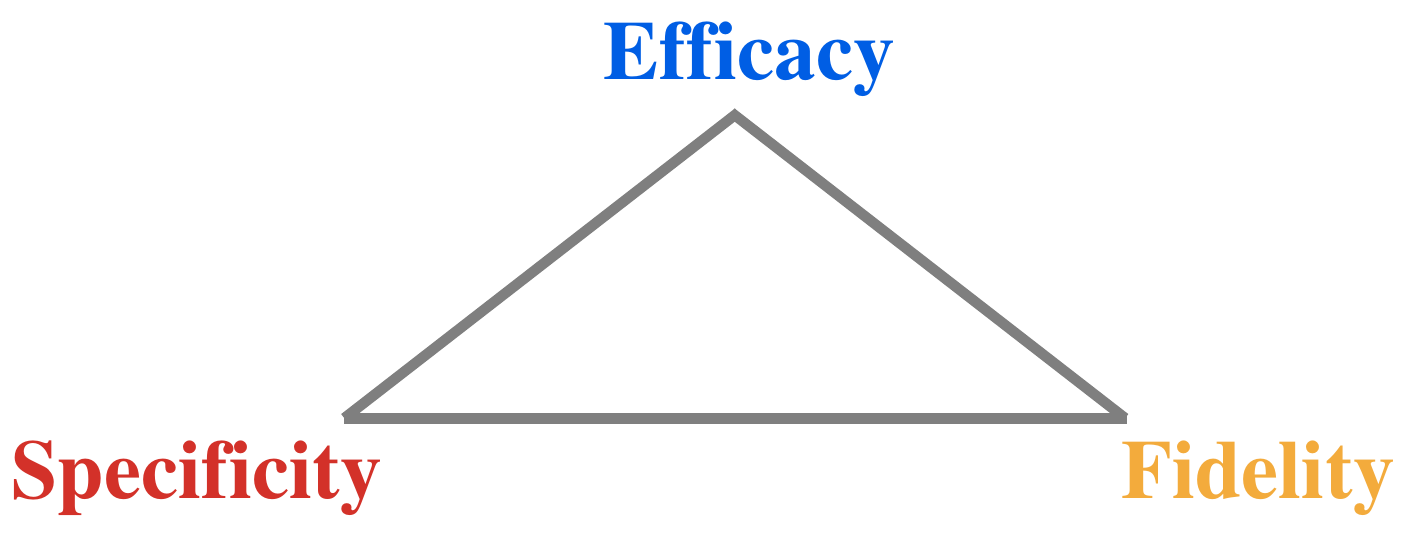}
\caption{The three desiderata of an adversary's attack~\cite{pang2020tale}. It was found that it is impossible for adversary's attacks to simultaneously and strictly achieve all the three desiderata of efficacy, fidelity, and specificity.}
\label{fig:motivation_efs_triagnle}
\end{figure}
}

In this section, we analyze the existing state-of-the-art backdoor attacks as well as their limitations which motivate our proposed backdoor attack residing in the frequency domain of images.
In particular, we analyze the existing backdoor attacks using the three desiderata of an adversary's attack~\cite{pang2020tale}, \ie {efficacy}, {specificity}, and {fidelity}.\footnote{Although it is controversial whether the poisoning image should keep high fidelity, it is definitely undesirable if an attack become suspicious or can be easily detected by a simple manual inspection from the adversary's perspective.} While an ideal backdoor attack should satisfy all the three aspects, it was found that it is impossible for adversary’s attacks to simultaneously and strictly achieve all the them~\cite{pang2020tale}.




\subsection{Existing Backdoor Attacks}

\begin{figure}[!t]
\centering
\subfigure[]{
  \label{fig:motivation_backdoor_attacks_a}
  \includegraphics[width=0.95\linewidth]{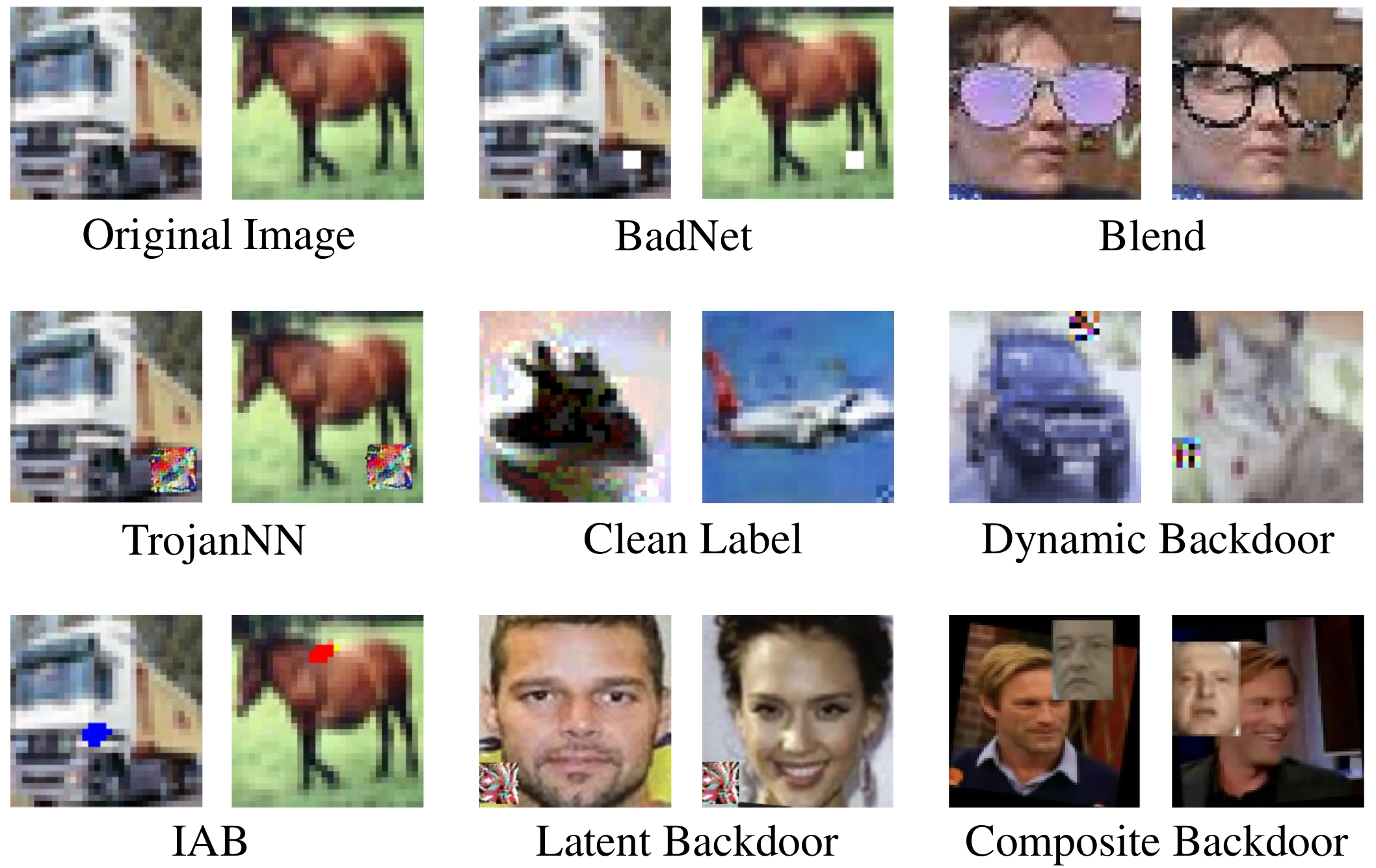}}
\subfigure[]{
  \label{fig:motivation_backdoor_attacks_b}
  \includegraphics[width=0.95\linewidth]{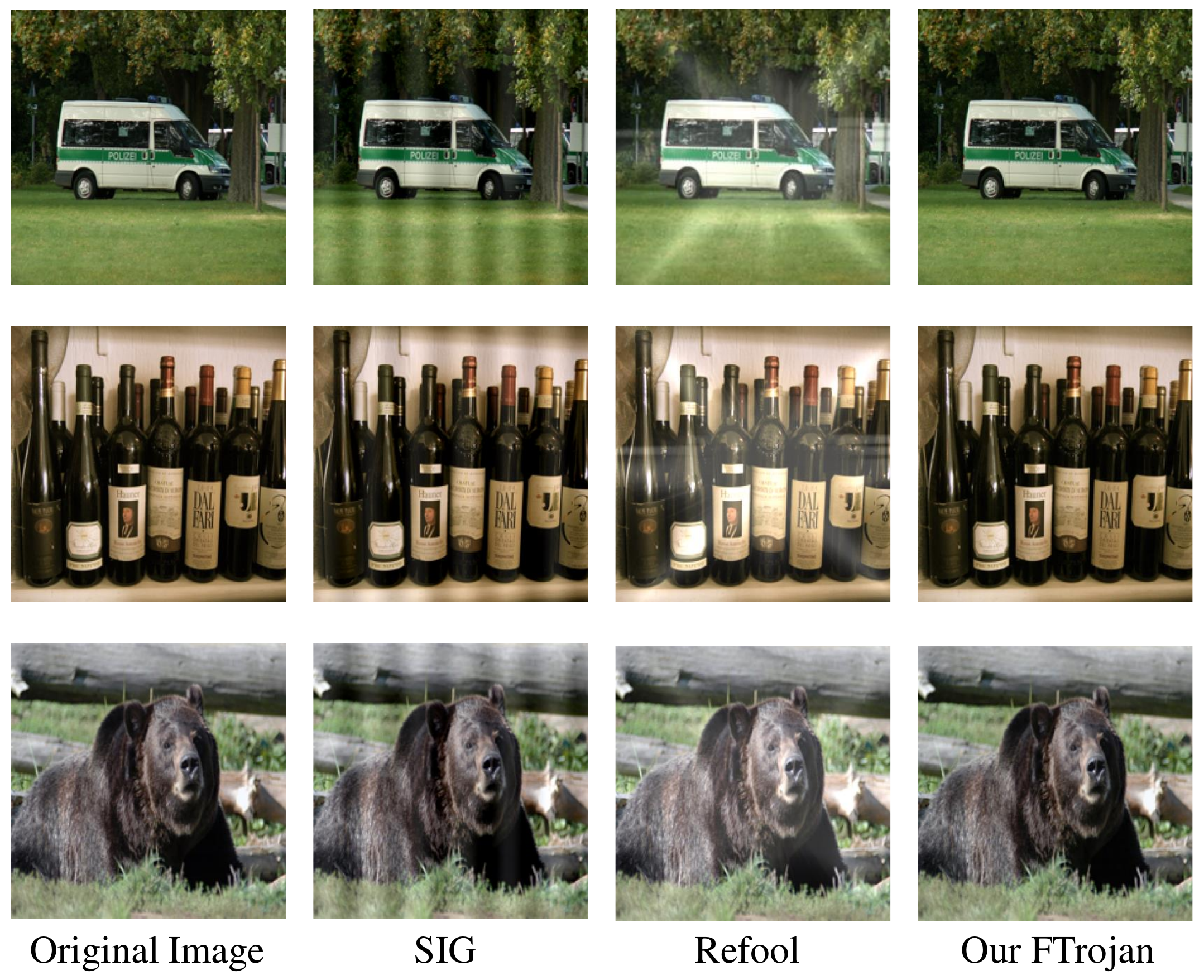}}
  \vspace{-1ex}
\caption{The poisoning images of existing backdoor attacks. (a) The example poisoning images from \badnet~\cite{gu2017badnets}, Blend~\cite{chen2017targeted}, TrojanNN~\cite{liu2018trojaning}, Clean Label~\cite{turner2018clean}, Dynamic Backdoor~\cite{salem2020dynamic}, \iab~\cite{nguyen2020input}, Latent Backdoor~\cite{yao2019latent}, and Composite Backdoor~\cite{lin2020composite}. Most of the images are directly copied from the original papers except for \badnet and \iab. As we can observe from the images, all these attacks are visually identifiable to a large extent. (b) The example poisoning images from \sig~\cite{barni2019new} and \refool~\cite{liu2020reflection}. Each column corresponds to one attack. Although the poisoning images are more natural, they are still detectable by humans due to their wave patterns in the background and the abnormal reflective phenomenon, respectively.}
\label{fig:motivation_backdoor_attacks}
\end{figure}

As a typical and pioneering backdoor attack, \badnet~\cite{gu2017badnets} proposes to poison some training data with predefined triggers (\eg a square in Figure~\ref{fig:motivation_backdoor_attacks_a}) for a target label. Then, it changes the labels of the poisoning images to the target label, and trains the CNN model with the poisoning data. For a test input stamped with the trigger, no matter what the real label is, the trained model will predict the test input as the target label. 
Later backdoor attacks mainly focus on making the attacks more effective against existing backdoor defenses. Typical strategies include generating different triggers for different inputs~\cite{salem2020dynamic,nguyen2020input} and reusing existing objects in the target label as triggers~\cite{lin2020composite}.
Recently, several backdoor attacks pay special attention to the fidelity aspect, by dispersing the trigger to a much larger area~\cite{barni2019new,liu2020reflection}. Consequently, the triggers are thus less visible to humans (see Figure~\ref{fig:motivation_backdoor_attacks_b} for some examples) and less detectable to existing defenses.
For example,
The \sig~\cite{barni2019new} attack transfers the images with superimpose signals (\eg a ramp signal or a sinusoidal signal), and triggers are contained in the varying background; the \refool~\cite{liu2020reflection} attack defines triggers resembling to the natural reflection phenomenon, and shows that it is resistant to several defenses including Fine-pruning~\cite{liu2018fine} and \nc~\cite{wang2019neural}.

{\em Limitations: the existing backdoor attacks are still largely visible and lack fidelity}. 
While the existing backdoor attacks perform relatively well on the efficacy and specificity aspects, they tend to fall short in terms of satisfy the fidelity requirement. Some poisoning images of the above attacks are illustrated in Figure~\ref{fig:motivation_backdoor_attacks}. Figure~\ref{fig:motivation_backdoor_attacks_a} shows the poisoning images of the backdoor attacks whose triggers are concentrated in a small area. As we can see, all the triggers are visually identifiable to a large extent, making the poisoning data easily detectable by humans. 
For \sig and \refool as shown in Figure~\ref{fig:motivation_backdoor_attacks_b}, although the triggers are dispersed to a larger area and thus more negligible compared to previous attacks, they are still generally detectable by humans (\eg the wave pattern in the background or the abnormal reflective phenomenon). Based on the low fidelity results of existing backdoor attacks, we speculate that the fundamental reason lies in that they directly inject or search for triggers in the spatial domain of an image. Therefore,
the triggers usually need to be sufficiently prominent, and thus potentially visible, so as to make CNN models recognize and remember their features.








\subsection{Our Key Insight}
In this work, we propose to launch the backdoor attack in the frequency domain. Our key insights are as follows. 
\begin{itemize}
    \item {\em First, attacks in the frequency domain can result in poisoning images with high fidelity}. On the one hand, existing backdoor attacks have shown that dispersing the trigger to the entire image in the spatial domain could improve the image fidelity and make the trigger less visible and more robust against existing defenses. On the other hand, given an image, a small perturbation in its frequency domain usually corresponds to tiny perturbations of a relatively large area in the spatial domain. 
    \item {\em Second, triggers in the frequency domain are recognizable and learnable by CNNs}. To implant an effective backdoor, we need to ensure that the trigger is learnable and can be memorized by CNNs, especially considering the fact that the trigger energy from frequency domain is dispersed throughout the entire image. To this end, it was recently observed that CNNs can recognize and memorize the features in the frequency domain of images due to the convolution operator~\cite{yin2019fourier,xu2019training,xu2019frequency,wang2020high}.
\end{itemize} 

\begin{figure}[t]
\centering
  \includegraphics[width=0.75\linewidth]{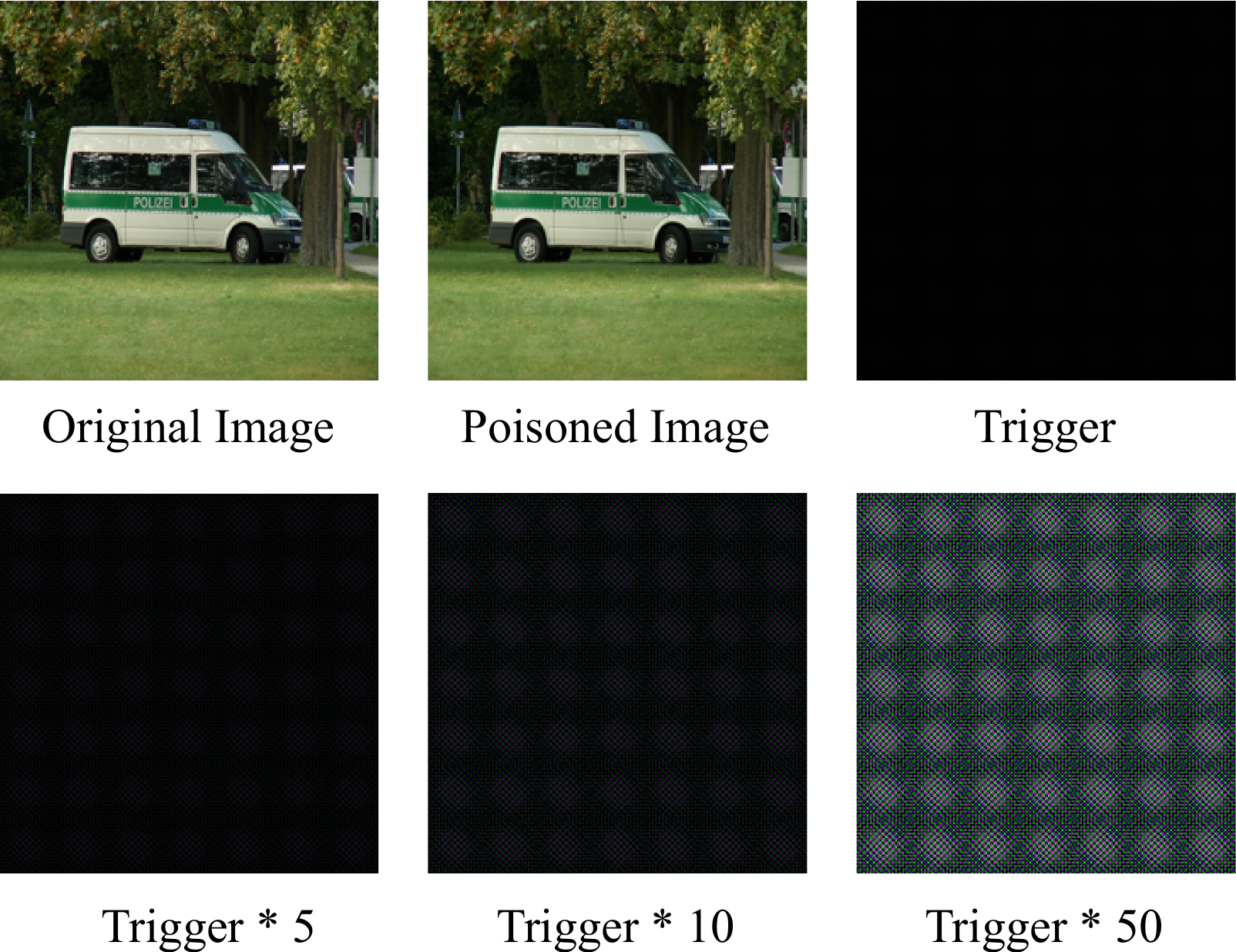}
\caption{An illustration of the trigger of \tool. The trigger is scattered over the entire image and nearly negligible to the HVS. To better show how the trigger looks like, we multiply each pixel value of the trigger with a given factor in the second row.}
\label{fig:motivation_frequency2spatial}
\end{figure}

Combining the above two insights together, we could inject a trigger (\eg a small perturbation on a high frequency) from the frequency domain, which is nearly invisible in the spatial domain but still learnable by CNNs.
An example of our trigger is shown in Figure~\ref{fig:motivation_frequency2spatial}. 
We can visually observe \hide{from Figure~\ref{fig:motivation_frequency2spatial} (and Figure~\ref{fig:motivation_backdoor_attacks_b} as well)} that the poisoning images by our method retain very high perceptual similarity to their original images. Additionally, we can observe from the first row of Figure~\ref{fig:motivation_frequency2spatial} that, the injected trigger is nearly invisible to humans.
To further show how the trigger looks like, we multiply each pixel value of the trigger by a factor and show the results in the second row of the figure. We can observe that the trigger is scattered over the entire image.

\begin{figure*}[t]
\centering
\includegraphics[width=0.85\linewidth]{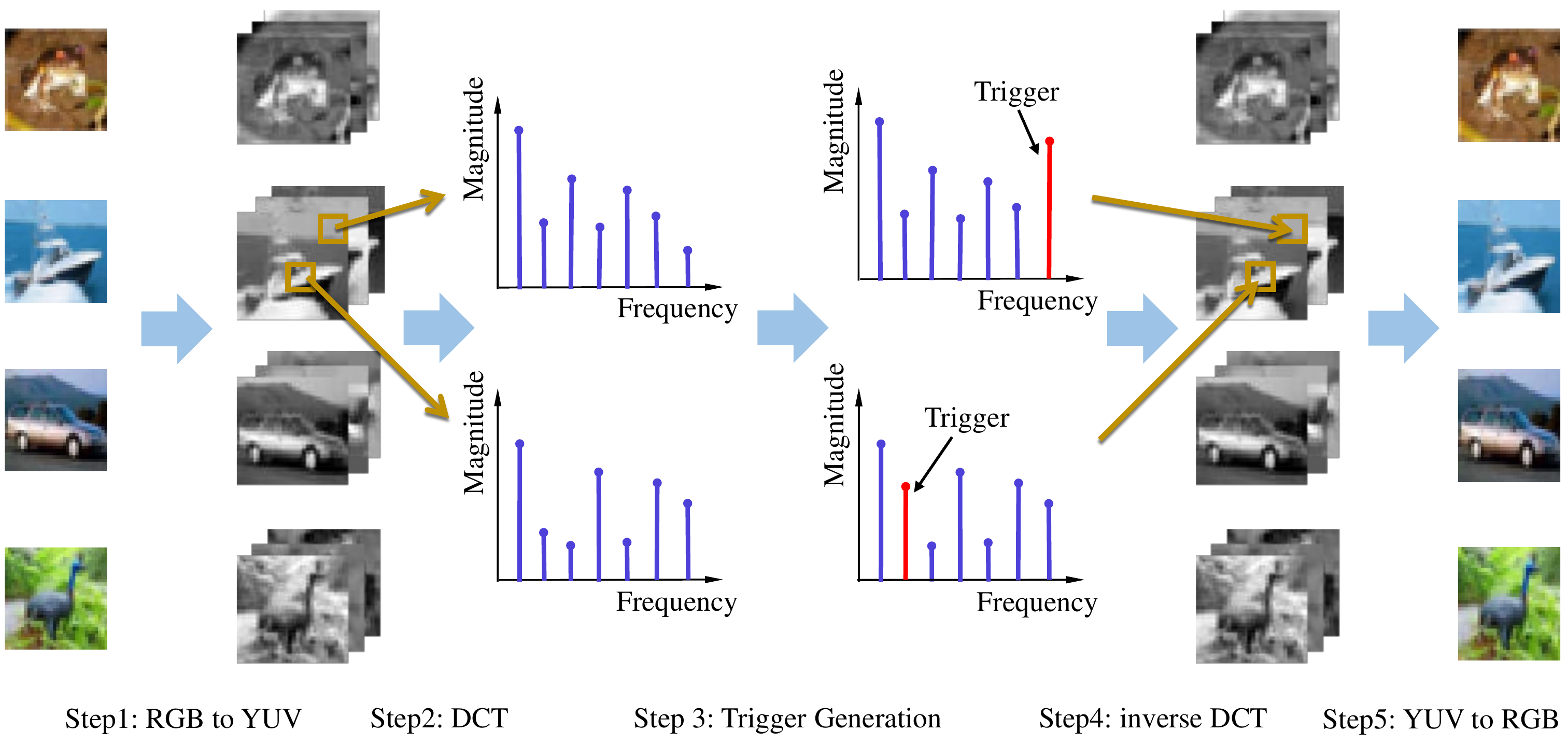}
 \vspace{-1ex}
\caption{The overview of the proposed backdoor attack \tool. After transforming the images to YUV channels and further to frequency domain, it places the trigger in the frequency domain and then transforms the images back to the spatial domain.}
\label{fig:design_pipeline}
\end{figure*}

\subsection{Threat Model}

We assume the adversary can access part of the training data. Such an assumption is the same with previous attacks~\cite{gu2017badnets,salem2020dynamic} and is practical when the training process is outsourced or the users have collected some training data from open repositories polluted by the adversary.
The adversary does not necessarily have the access to the parameters of the CNN model.
The adversary's goal is to make the backdoored model misclassify inputs that contain triggers to the target label, while behaving normally for benign inputs.

\section{Attack Design}\label{sec:design}

\subsection{Overview}

Figure~\ref{fig:design_pipeline} shows the poisoning process of the proposed backdoor attack \tool. Specifically, it consists of the following five steps.

{\em Step 1: color channel transform from RGB to YUV}. Given an input RGB image, we first convert it to YUV channels. 
The reason is that YUV channels contain the bandwidth for chrominance components (\ie UV channels) that are less sensitive to the HVS. Therefore, injecting triggers in the chrominance components could be more negligible to prevent human perception. 

{\em Step 2: discrete cosine transform from spatial domain to frequency domain}. 
We next transform the UV channels of the image from the spatial domain to the frequency domain via discrete cosine transform (DCT). Here, a small perturbation on the frequency domain may correspond to a large area in the spatial domain. 

{\em Step 3: trigger generation in the frequency domain}. \tool chooses a {\em frequency band} with a fixed magnitude in the frequency domain to serve as the trigger. 
In particular, we consider trigger generation strategies related to what frequency is the trigger placed on and what is the magnitude of the trigger. 

{\em Step 4: inverse discrete cosine transform from frequency domain to spatial domain}. After the frequency trigger is generated, we apply inverse DCT to obtain the poisoning image in the spatial domain denoted by YUV channels.

{\em Step 5: color channel transform from YUV to RGB}. Finally, since CNN models are mainly trained on the RGB color space, we transform the YUV channels back to the RGB channels.

\hide{Finally, we follow the existing backdoor attacks by poisoning a small subset of images and changing the label of the poisoning images to the target label. We then use the modified training set to train the CNN model.}
Note that, once the trigger is defined in the frequency domain, it corresponds to fixed pixels (with fixed values) in the spatial domain. Therefore, we can use these pixels as the trigger to superimpose the original pixels to poison an image, without the need of repeatedly computing the above transforms. This means that our attack can be used in real-time scenarios. In the physical world, we can directly superimpose the trigger pixels into test images by, \eg pasting a near-transparent film containing only trigger pixels.
\subsection{Color Channel Transform}

RGB is the most commonly used color space for computer screens and CNN models. The contributions of the three channels are equal to the visual perception of the image. In contrast, the YUV space divides a color image into luminance components (\ie Y channel) and chrominance components (\ie U and V channels), and human visual perception is less sensitive to the latter.
Therefore, although we can directly inject triggers in the RGB space, we choose to do so in the UV channels for better image fidelity. Additionally, perturbations in UV channels could affect all the three RGB channels, making CNNs trained on RGB channels easier to recognize their features. We will study different design choices here in the experimental evaluations. The transform equations are shown in the appendix (Appendix~\ref{app:yuv}) for completeness.




\begin{figure}[t]
\centering
\includegraphics[width=0.55\linewidth]{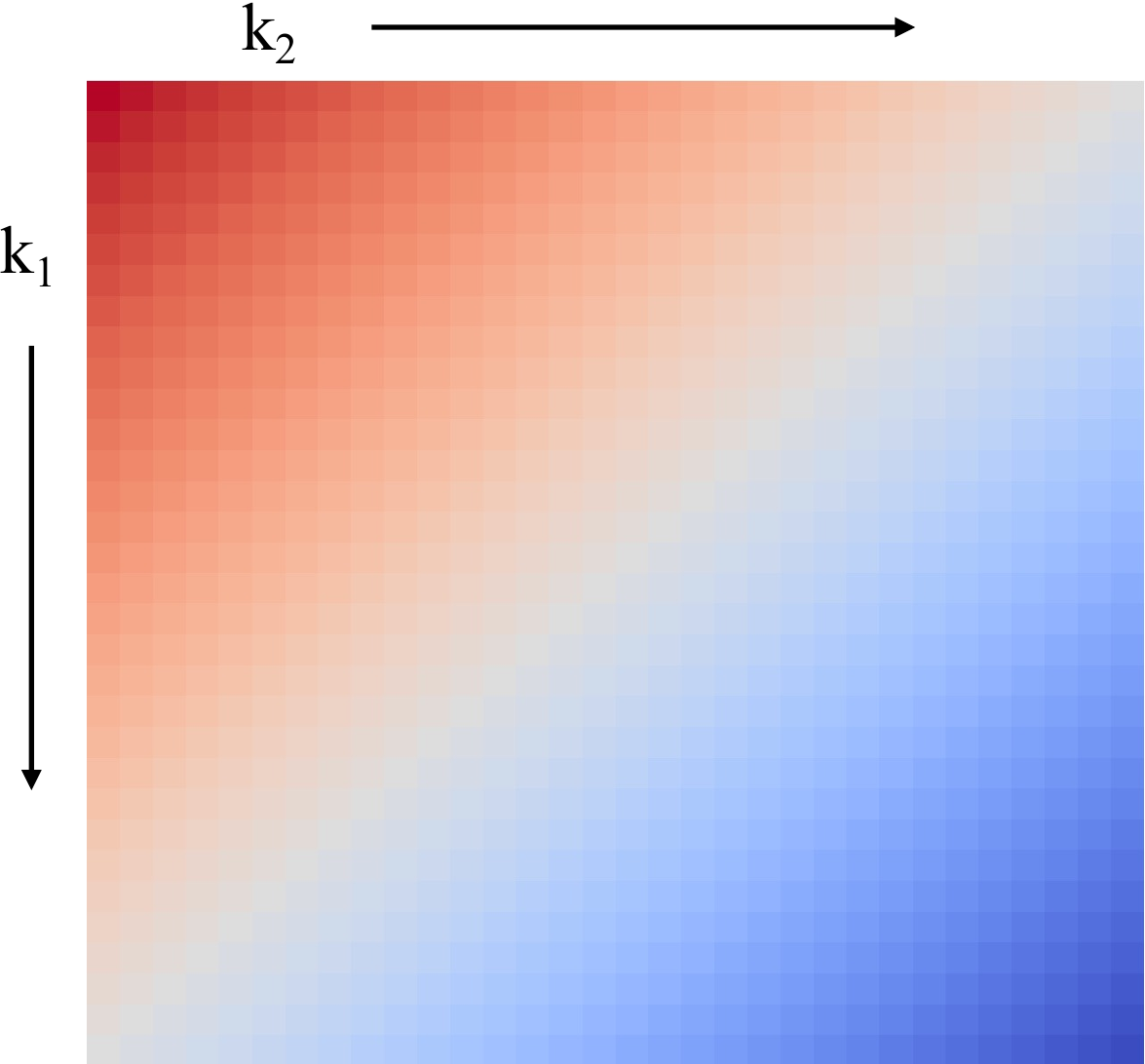}
\caption{Frequency map of DCT. Each frequency band is indicated by the 2-D frequency index ($k_1, k_2$). The frequency goes from low to high along the diagonal direction from top left to bottom right.}
\label{fig:design_frequency_map}
\end{figure}


\subsection{Discrete Cosine Transform}
Given a channel that we aim to inject the trigger, we next transform the channel from spatial domain to frequency domain. In particular, we choose to use DCT which expresses an image as a set of cosine functions oscillating at different frequencies. Compared with discrete Fourier transform (DFT), DCT is better in terms of energy concentration and is widely used in processing images. 
An example frequency map of DCT is shown in Figure~\ref{fig:design_frequency_map}, which contains $32 \times 32$ {frequency bands} transformed from an image of size $32 \times 32$.
After the image is transformed by DCT, most of its energy is concentrated in the low-frequency component (\ie in the upper left corner of the frequency map indicated by dark brown), and the high-frequency component is in the bottom right corner (indicated by dark blue). We name the areas close to the white diagonal (from the bottom left to the top right) as the mid-frequency component.
In the following, we use {\em index $(k_1, k_2)$} to indicate the frequency band as shown in Figure~\ref{fig:design_frequency_map}.


In this work, we use the 2-D Type-II DCT transform~\cite{ahmed1974discrete} as follows,
\begin{eqnarray}\label{eq:DCT}
X(k_1, k_2) &=& \sum_{n_1=0}^{N_1-1}\sum_{n_2=0}^{N_2-1}x(n_1,n_2)c_1(n_1,k_1)c_2(n_2,k_2), \nonumber\\
c_i(n_i,k_i)&=& \widetilde{c}_i(k_i)\cos(\frac{\pi (2n_i+1)k_i}{2N_i}), \nonumber\\
\widetilde{c}_i(k_i) &=&
\begin{cases}
\frac{1}{\sqrt{N_i}}& k_i=0\\
\frac{2}{\sqrt{N_i}}& k_i \neq 0
\end{cases} \;\; i=1, 2,  
\end{eqnarray}
which transforms the size $N_1 \times N_2$ input image in spatial domain to its frequency domain with the same size. Here, $(k_1, k_2)$ stands for the index as shown in Figure~\ref{fig:design_frequency_map}, $k_1/k_2 \in \{0, 1, \dots, N_1/N_2\}$, $X(k_1, k_2)$ is the frequency magnitude at $(k_1, k_2)$, and $x(n_1, n_2)$ is the pixel value in position $(n_1, n_2)$ of the image in spatial domain. To transform the image from frequency domain back to spatial domain, we can use the inverse DCT transform~\cite{rao2014discrete} whose equations are similar to Eq.~\eqref{eq:DCT} and thus omitted for brevity.

In practice, we divide the images into a set of disjoint {\em blocks} and perform DCT on each block. Blocks that are too large would make the computation time-consuming, and too small could cause serious distortion to the image. We set $N1 = N2 = 32$ in this work. For an image whose size is larger than $32 \times 32$, we poison all the blocks by default as this strategy still results in high-fidelity images. Other design choices such as choosing to poison smaller blocks or fewer blocks are shown in Appendix~\ref{app:moreexp}.

\subsection{Trigger Generation}\label{sec:trigger_generation}
Trigger generation involves the following two orthogonal dimensions, \ie {\em trigger frequency} and {\em trigger magnitude}. 

\eat{
\begin{figure}[t]
\begin{center}
\includegraphics[width=0.9\linewidth]{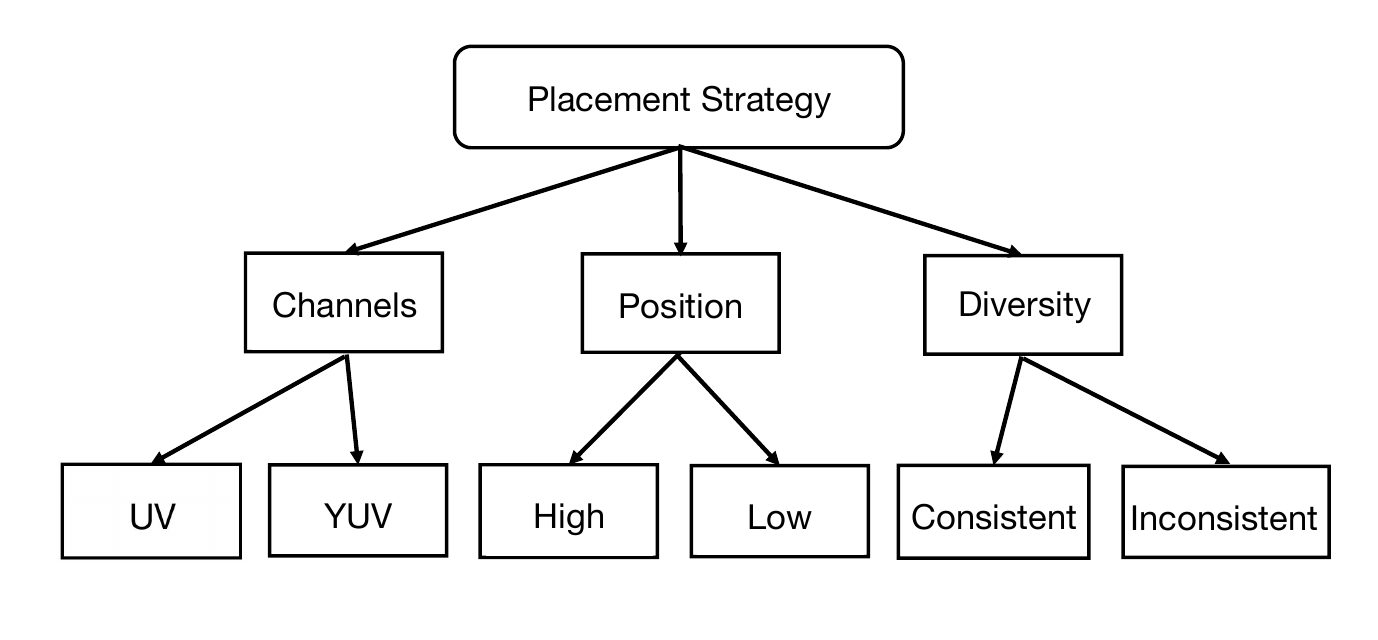}
\end{center}
\caption{Trigger Frequency Placement Strategy}
\label{fig:design_strategy}
\end{figure}
}

\begin{figure*}[t]
\centering
\includegraphics[width=0.92\linewidth]{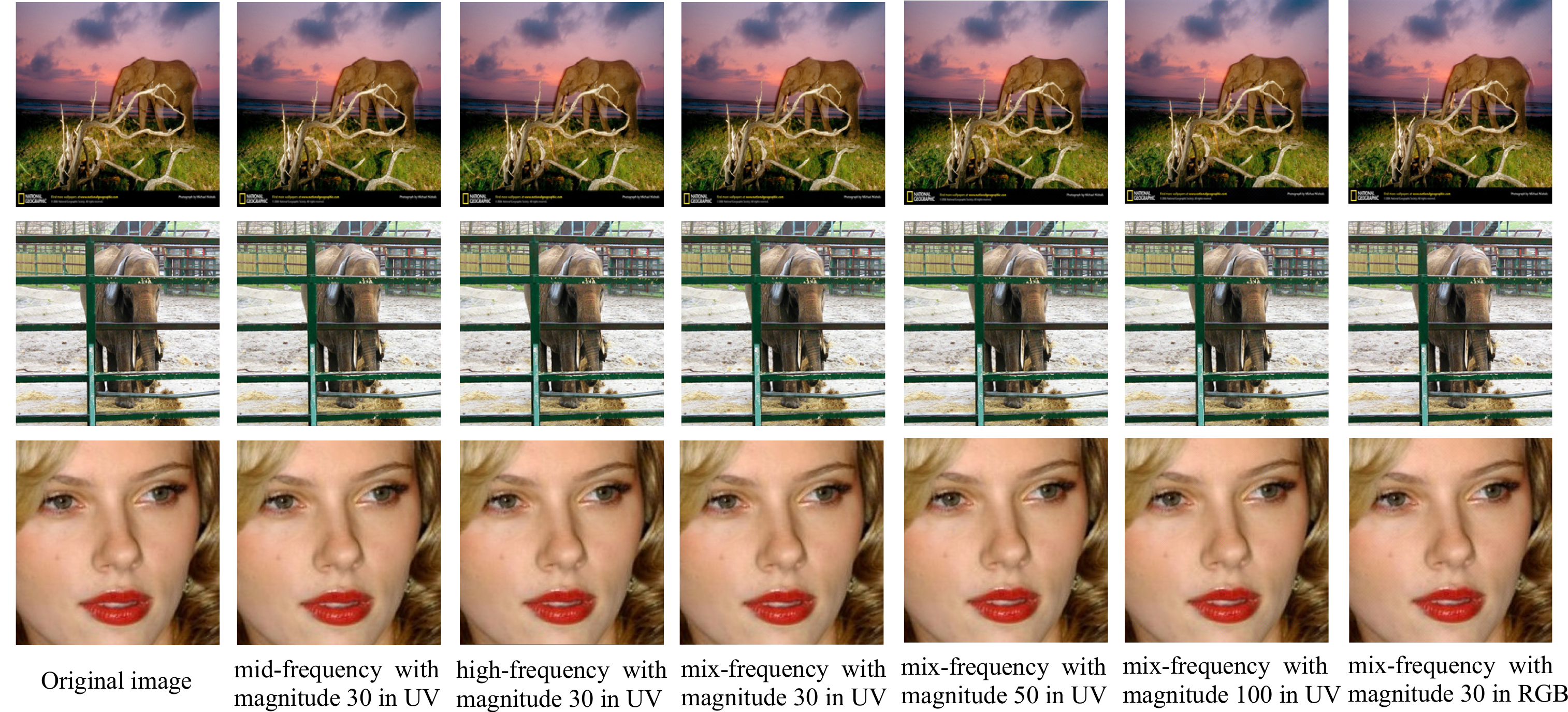}
 \vspace{-1ex}
\caption{Poisoning images by our \tool attack. Mix-frequency mixes triggers in both mid- and high-frequency components. We can observe that when the triggers reside in the high-frequency and mid-frequency components with moderate magnitude, the poisoning images are difficult to visually detect.}
\label{fig:design_attack_image}
\end{figure*}

{\em {Trigger frequency}}. First, we need to decide the specific frequency band that we aim to place the trigger on. 
On the one hand, placing the trigger at higher frequency would make the poisoning image less sensitive to human perception, but such triggers could be erased by low-pass filters. On the other hand, triggers at lower frequency is robust against low-pass filters but could cause visual abnormalities if the trigger magnitude is too large. In this work, we choose a more robust {\em mix mode}, \ie placing one trigger at mid frequency and one at high frequency for each block.

{\em {Trigger magnitude}}. In general, larger magnitude may be easier for CNNs to learn and also robust against some low-pass filters; however, it also comes at a risk of being detected by human perception or existing backdoor defenses. Smaller magnitude may bypass human perception and existing defenses, but being attenuated by the low-pass filters. We evaluate different choices in the experiment and choose a moderate magnitude depending on the specific datasets.




Figure~\ref{fig:design_attack_image} shows the poisoning images with different trigger frequencies and trigger magnitudes. All the images are stamped with triggers in UV channels of each block.
The original clean images are in the first column, and the rest columns contain the poisoning images.  The fifth column stands for our default setting, with triggers in mix mode (indexed by $(15, 15)$ and $(31, 31)$) of magnitude 50. 
We can observe that when the triggers reside in either mid-frequency or high-frequency bands with moderate magnitude (\eg no more than 100), the poisoning images are perceptually similar to the corresponding clean images and difficult to visually detect. 

\section{Evaluation}\label{sec:exp}
\begin{table*}[t]
\centering
\caption{Summary of the datasets and the classifiers used in our experiments.}
\label{table:eval_summary_of_dataset}
\resizebox{1\textwidth}{!}{
\begin{tabular}{llrrrc}
\toprule
Task & Dataset & \# of Training/Test Images & \# of Labels & Image Size  & Model Architecture \\
\midrule   
  Handwritten Digit Recognition & MNIST & 60,000/10,000 & 10  &  $32 \times 32 \times 1 $ & 2 Conv + 2 Dense   \\
  Traffic Sign Recognition & GTSRB & 39,209/12630 & 43 &  $32\times 32 \times 3$  & 6 Conv + 1 Dense \\   
  Object Classification & CIFAR10 & 50,000/10,000 & 10 & $32\times 32\times 3$  & 6 Conv + 1 Dense \\   
  Object Classification & ImageNet & 20,567/1,315 & 16 &  $224 \times 224 \times 3$ & ResNet50 \\
  Face Recognition & PubFig	& 5,274/800  & 60 &  $224\times 224 \times 3$ & ResNet50 \\   
  \bottomrule
\end{tabular}
}
\end{table*}

In this section, we evaluate our backdoor attack \tool in terms of the following two dimensions:
\begin{itemize}
    \item The efficacy, specificity, and fidelity aspects of the proposed attack, with different design choices and comparisons to existing attacks.
    \item The robustness of the proposed attack under defending techniques.
\end{itemize}

\subsection{Experimental Setup}

\begin{table*}
  \centering
  \caption{Efficacy and specificity results of \tool variants. All the results are percentiles. For the default \tool (\ie `ABSF+UV+mix' variant), it can achieve 98.78\% ASR, while the BA decreases by 0.56\% on average.}
  \label{table:eval_basic_experiment1}
  \begin{tabular}{crrrrrrrrrrrrrrr}
   \toprule
   \multirow{2}{*}{\tool Variant}  & \multicolumn{2}{c}{MNIST} & \multicolumn{2}{c}{GTSRB} & \multicolumn{2}{c}{CIFAR10} & \multicolumn{2}{c}{ImageNet} & \multicolumn{2}{c}{PubFig} \\
   \cmidrule(lr){2-3} \cmidrule(lr){4-5} \cmidrule(lr){6-7} \cmidrule(lr){8-9} \cmidrule(lr){10-11} & BA & ASR & BA & ASR & BA & ASR  & BA & ASR  & BA & ASR  \\
   \midrule
   
No attack & 99.40 & - & 97.20 & - & 87.12 & - & 79.60 & - & 89.50 & - \\\midrule

UV+mix & 99.36 & 99.94 & 96.63 & 99.25 & 86.05 & 99.97 & 78.63 & 99.38 & 88.62 & 99.83 \\
UV+mid & 99.40 & 99.22 & 96.91 & 98.59 & 86.90 & 99.90 & 78.50 & 99.75 & 89.13 & 97.86 \\
UV+high & 99.39 & 99.81 & 96.63 & 99.12 & 86.90 & 99.90 & 78.75 & 99.14 & 88.25 & 99.93 \\
YUV+mix & - & - & 96.82 & 98.35 & 86.76 & 99.96 & 79.13 & 99.38 & 88.08 & 99.93 \\
RGB+mix & - & - & 97.16 & 92.05 & 86.33 & 95.99 & 78.70 & 95.46 & 89.37 & 99.25 \\


    
    \bottomrule
\end{tabular}
\end{table*}

\subsubsection{Tasks, Datasets, and Models}
We conduct experiments on several benchmark tasks/datasets,
with details described below and summarized in Table~\ref{table:eval_summary_of_dataset}.
\begin{itemize}
\item {\em Handwritten digit recognition on MNIST data}~\cite{lecun1998gradient}. The MNIST dataset contains 60,000 training samples and 10,000 test samples of handwritten digits (0 - 9). For this dataset, we train a CNN model with two convolutional layers and two fully-connected layers. The goal is to identify 10 handwritten digits from the grayscale images. Before sending the training data to the model for training, we made padding to resize the image from $28 \times 28$ to $32 \times 32$. 
\item {\em Traffic sign recognition on GTSRB data}~\cite{stallkamp2011german}. GTSRB contains 43 different traffic signs simulating the application scenario of autonomous driving. There are 39,209 color training images and 12,630 test images. When processing the GTSRB dataset, we found that these images are very different in illumination. Thus, we followed the instructions\footnote{https://benchmark.ini.rub.de/gtsrb\_dataset.html\#Pre-calculated\_feature} and performed histogram equalization on these images in the HSV color space. We also adjusted these images to the same standard size.

\item {\em Object classification on CIFAR10 data}~\cite{krizhevsky2009learning}. The CIFAR10 dataset contains 60,000 color images of size $32 \times 32$ in 10 different classes. We split it into 50,000 training images and 10,000 test images. For both GTSRB and CIFAR10 images, since they are more complicated compared to the images in MNIST, we train a CNN with six convolutional layers and one fully-connected layer.

\item {\em Object classification on ImageNet data}~\cite{deng2009imagenet}. ImageNet is also an object classficiation dataset but with higher resolution. In particular, we randomly sampled 16 labels in the original ImageNet dataset, and then split the images into 20,567 training images and 1,315 test images.

\item {\em Face recognition on PubFig data}~\cite{krizhevsky2009learning}. PubFig is a real-world dataset consisting of face images of 200 persons collected from the Internet. We use the sampled subset of 60 persons from~\cite{liu2020reflection}. The dataset contains 5,274 training images and 800 test images. For both ImageNet and Pubfig, we resize all the images to size $224 \times 224 \times 3$ and train a ResNet50 model~\cite{he2016deep}.
\end{itemize}

\small
\begin{table*}
  \centering
  \caption{Fidelity results of \tool variants. Larger PSNR and SSIM, and smaller IS are better. Triggering at UV channels achieves better results than at YUV and RGB channels. Triggering at high-frequency only is slightly better than at mid-frequency and mix-frequency.}
  \label{table:eval_basic_experiment2}
  \resizebox{1\textwidth}{!}{
  \begin{tabular}{crrrrrrrrrrrrrrr}
   \toprule
   \multirow{2}{*}{\tool Variant}  & \multicolumn{3}{c}{GTSRB} & \multicolumn{3}{c}{CIFAR10} & \multicolumn{3}{c}{ImageNet} & \multicolumn{3}{c}{PubFig} \\
   \cmidrule(lr){2-4} \cmidrule(lr){5-7} \cmidrule(lr){8-10} \cmidrule(lr){11-13} & PSNR & SSIM & IS & PSNR & SSIM & IS & PSNR & SSIM & IS & PSNR & SSIM & IS \\
   \midrule
   
No Attack & INF & 1.000 & 0.000 & INF & 1.000 & 0.000 & INF & 1.000 & 0.000 & INF & 1.000 & 0.000 \\\midrule
UV+mix & 40.9 & 0.995 & 0.017 & 40.9 & 0.995 & 0.135 & 37.7 & 0.727 & 0.020 & 37.7 & 0.802 & 0.213 \\
UV+mid & 43.3 & 0.995 & 0.011 & 43.5 & 0.997 & 0.098 & 40.5 & 0.775 & 0.014 & 40.5 & 0.861 & 0.176 \\
UV+high & 43.3 & 0.995 & 0.007 & 43.5 & 0.997 & 0.049 & 40.5 & 0.796 & 0.009 & 40.5 & 0.870 & 0.019 \\
YUV+mix & 25.7 & 0.943 & 0.458 & 36.5 & 0.985 & 0.279 & 25.7 & 0.670 & 0.258 & 21.3 & 0.806 & 1.571 \\
RGB+mix & 45.8 & 0.995 & 0.012 & 45.7 & 0.997 & 0.046 & 40.4 & 0.784 & 0.045 & 41.3 & 0.861 & 0.282 \\


   \bottomrule
\end{tabular}
}
\end{table*}
\normalsize

\subsubsection{Evaluation Metrics}
For efficacy and specificity, we measure the {\em attack success rate (ASR)} and the {\em  accuracy on benign data (BA)}, respectively. For fidelity, it is still an open problem to measure it. In this work, we mainly consider if human eyes are sensitive to the poisoning images and use metrics {\em peak signal-to-noise ratio (PSNR)}~\cite{huynh2008scope}, {\em structural similarity index (SSIM)}~\cite{wang2004image}, and {\em inception score (IS)}~\cite{salimans2016improved,barratt2018note}.

\begin{itemize}
\item {\em ASR}. ASR defines the percentage of samples in the test set that are incorrectly and successfully classified as the target label after being injected with a trigger. Higher ASR indicates better efficacy.

\item {\em BA}. BA is the accuracy of the backdoored model on the clean test set. If the BA score is close to the accuracy of the clean model on the test set, it means that such an attack has a high specificity.

\item {\em PSNR}. As the name suggests, PSNR measures the ratio between the maximum pixel value of an image and the mean squared error (between clean and poisoning images). A larger PSNR means that the perceptual difference between the two images is smaller.

\item {\em SSIM}. SSIM is an index to measure the similarity of two images. It is calculated based on the luminance and contrast of local patterns. A larger SSIM indicates that the poisoning images are of better quality.

\item {\em IS}. IS is a widely-used metric to measure the perceptual quality of images generated from GANs. \eat{It mainly considers two aspects, \ie clarity and diversity of the generated images.} It uses features of the InceptionV3 network~\cite{szegedy2016rethinking} trained on ImageNet to mimic human visual perception. We compute the KL divergence between the features, and a smaller IS means better perceptual quality. 
\end{itemize}

\subsubsection{Implementations}
For the proposed \tool attack, we implement it with two versions in both PyTorch and Tensorflow 2.0, and the code is available at our project website~\cite{ftrojan}. Our default settings are as follows. For trigger frequency, we place the trigger at frequency bands $(15,15)$ and $(31,31)$ where $(15,15)$ belongs to the mid-frequency component and $(31, 31)$ belongs to the high-frequency component. Based on the size of images, we set the trigger magnitude to 30 for MNIST, CIFAR10, GTSRB, and 50 for ImageNet, PubFig. The injection rate is fixed to 5\% for simplicity. We use the Adam optimizer with learning rate 0.0005 for MNIST and GTSRB, and the RMSprop optimizer with learning rate 0.001 for the rest datasets. The batch size is set to 64. In the following, we use \tool to denote the default setting unless otherwise stated. The target label is set to 8 for all the datasets.
All the experiments were carried out on a server equipped with 256GB RAM, one 20-core Intel i9-10900KF CPU at 3.70GHz and one NVIDIA GeForce RTX 3090 GPU.

\subsection{Attack Performance}

\subsubsection{Overall Performance}

We first evaluate different trigger generation strategies of the proposed \tool attack. The efficacy and specificity results are shown in Table~\ref{table:eval_basic_experiment1}, and the corresponding fidelity results are shown in Table~\ref{table:eval_basic_experiment2}. 
For the variants in the table, 
`UV', `YUV', and `RGB' indicate injected channels of the trigger,\footnote{The MNIST images are gray-scale and have only one channel. We exclude the results on MNIST in Table~\ref{table:eval_basic_experiment2} and directly inject the trigger into this channel for Table~\ref{table:eval_basic_experiment1}.} and `mid', 'high', and `mix' mean the trigger frequencies. Here, `mix' is our default setting as mentioned above, and frequency bands $(15, 15)$ and $(31, 31)$ are used for `mid' and `high', respectively. 

There are several observations from the two tables. First, all the \tool variants are effective, namely, decreasing little on BA and having a high ASR. For example, on average, the default \tool (\ie `UV+mix') can achieve 98.78\% ASR, while the BA decreases by only 0.56\%.
Second, comparing different trigger frequencies, we can observe that all the three choices are closely effective and trojaning at high frequency tends to have higher fidelity results in general. 
Third, although as effective as the default \tool, injecting triggers into YUV channels instead of UV channels results in worse fidelity results as indicated by Table~\ref{table:eval_basic_experiment2} (the fifth row).
Fourth, injecting triggers at RGB channels is less effective than at UV channels, and it also results in lower fidelity (sixth row in Table~\ref{table:eval_basic_experiment2}). This is probably due to that the frequencies are more messy in RGB channels.

\begin{figure}[t]
\centering
\subfigure[GTSRB data]{
  \label{fig:eval_performance_vs_IR_GTSRB}
  \includegraphics[width=0.48\linewidth]{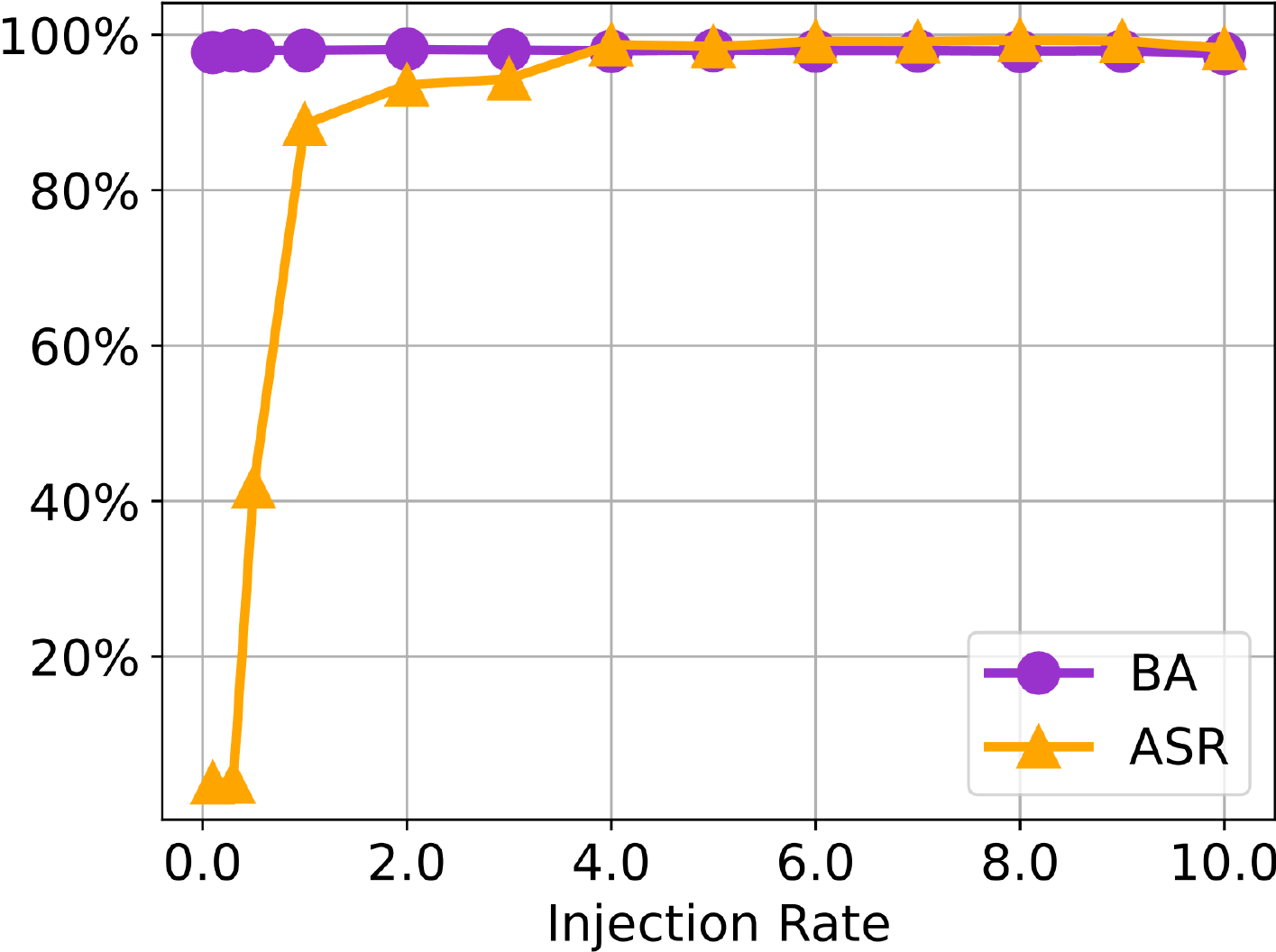}}
  \vspace{0.1in}
\subfigure[CIFAR10 data]{
  \label{fig:eval_performance_vs_IR_CIFAR10}
  \includegraphics[width=0.48\linewidth]{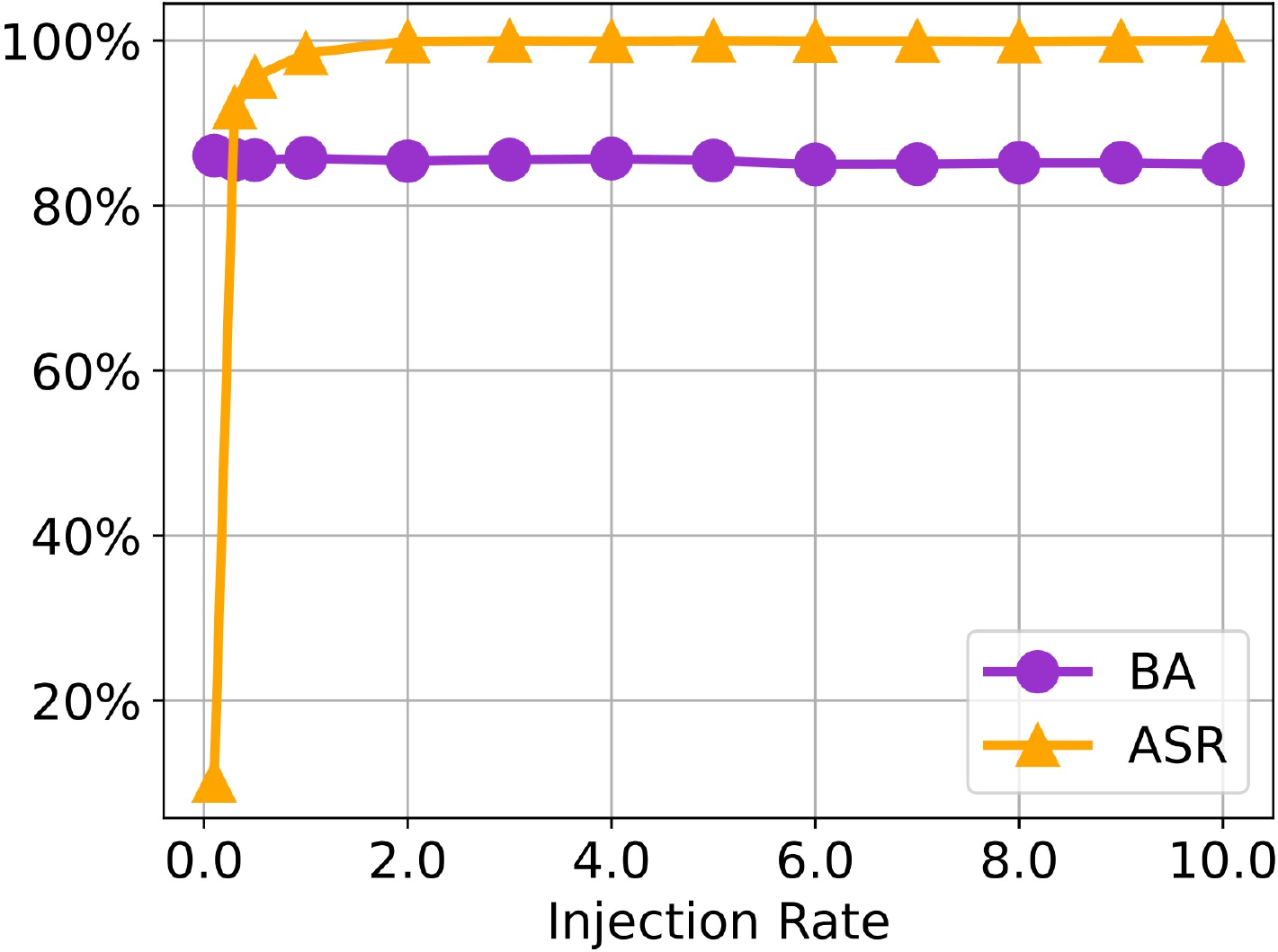}}
   \vspace{-1ex}
\caption{Performance vs. injection rate. \tool can achieve a high ASR when the injection rate is around 0.1\% - 1\%. We fix it to 5\% in this work to ensure high ASR.}
\label{fig:eval_performance_vs_IR}
\end{figure}

\subsubsection{Performance versus Injection Rate}
We next evaluate the effectiveness of \tool when the injection rate of poisoning images in training data varies. We increase the injection rate from 0.01\% to 10\% and show the results in Figure~\ref{fig:eval_performance_vs_IR}. In the following, we mainly report results on GTSRB and CIFAR10 as training on these two datasets is more efficient.
We can observe from the figure that BA does not change significantly when the injection rate is in a wide range. Additionally, when the injection rate is no less than 1\%, \tool can achieve a high ASR for both datasets. 
This experiment also shows that different datasets have different sensitivity to the injection rate. For example, injecting 0.1\% poisoning images could already achieve a high ASR on CIFAR10.

\begin{table}[t]
  \centering
  \caption{Performance vs. trggier frequency. All the results are percentiles. Triggering at mid- or high-frequency components generally results in better BA and ASR results.}
  \label{table:eval_frequency_position}
  \begin{tabular}{crrrrr}
   \toprule
   \multirow{2}{*}{Frequency Index}  & \multicolumn{2}{c}{GTSRB} & \multicolumn{2}{c}{CIFAR10} \\
   \cmidrule(lr){2-3}\cmidrule(lr){4-5} & BA & ASR & BA & ASR \\  
   \midrule
   (2, 6) & 94.60 & 81.49 & 84.12 & 84.16 \\
   (4, 4) & 94.79 & 44.71 & 84.59 & 70.85 \\
   (8, 8) & 95.77 & 77.11 & 82.79 & 13.72 \\
   (8, 20) & 96.11 & 94.63 & 85.49 & 96.91 \\
   (12, 12) & 97.11 & 96.65 & 86.44 & 90.36 \\
   (12, 16) & 96.69 & 91.75 & 86.95 & 99.36 \\
   (20, 20) & 96.60 & 95.21 & 85.95 & 99.71 \\
   (24, 24) & 96.62 & 94.27 & 86.76 & 99.58 \\
   (28, 28) & 96.66 & 98.73 & 86.95 & 99.94 \\
   \bottomrule
\end{tabular}
\end{table}

\begin{table*}
  \centering
  \caption{Comparison results with existing attacks. All the BA and ASR results are percentiles. Larger PSNR and SSIM, and smaller IS are better. \tool achieves higher ASR than the competitors on both datasets, and it outperforms the competitors for all the three fidelity metrics. Best results are in bold.}
  \label{table:eval_comparison_experiment}
  \begin{tabular}{crrrrrrrrrrrrrr}
  \toprule
   \multirow{2}{*}{Attack Method}  & \multicolumn{5}{c}{GTSRB} & \multicolumn{5}{c}{CIFAR10} \\
   \cmidrule(lr){2-6}\cmidrule(lr){7-11} & BA & ASR & PSNR & SSIM  &  IS & BA & ASR & PSNR & SSIM & IS \\ \midrule
   No Attack & 97.20 & - & INF & 1.000 & 0.000 & 87.12 & - & INF & 1.000 & 0.000  \\\midrule
    \badnet & 96.51 & 84.98 & 24.9 & 0.974 & 0.090 & 86.01 & 94.80 & 23.8 & 0.941 & 0.149  \\
    \sig & 96.49 & 92.56 & 25.3 & 0.973 & 1.353 & 85.70 & 95.76 & 25.2 & 0.871 & 1.905  \\
    \refool & 96.41 & 56.52 & 19.1 & 0.923 & 1.035 & 85.87 & 73.20 & 17.3 & 0.769 & 0.910 \\
    \iab & 92.12 & 64.84 & 23.8 & 0.956 & 0.226 & 85.10 & 79.70 & 13.2 & 0.829 & 2.240 \\
    
    \tool & {\bf 96.63} & {\bf 99.25} & {\bf 40.9} & {\bf 0.995} & {\bf 0.017} & {\bf 86.05} & {\bf 99.97} & {\bf 40.9} & {\bf 0.995} & {\bf 0.135} \\
\bottomrule
\end{tabular}
\end{table*}

\subsubsection{Performance versus Trigger Frequency}

For trigger frequency, we study different frequency indices while keeping the other settings as default. Specifically, we place the trigger on several low-frequency (\ie (4,4), (8,8), (8,16)), mid-frequency (\ie (8, 20), (12, 12), (12, 16)), and high-frequency (\ie (20, 20), (24, 24), (28, 28)) components, and the results are shown in Table~\ref{table:eval_frequency_position}. It can be seen that the backdoor attack is effective when the triggers are placed on mid- and high-frequency components. In this work, we choose a mix mode by default, \ie triggering one mid-frequency index and one high-frequency index.


\begin{figure}[t]
\centering
\subfigure[GTSRB data]{
  \label{fig:eval_magnitude_GTSRB}
  \includegraphics[width=0.48\linewidth]{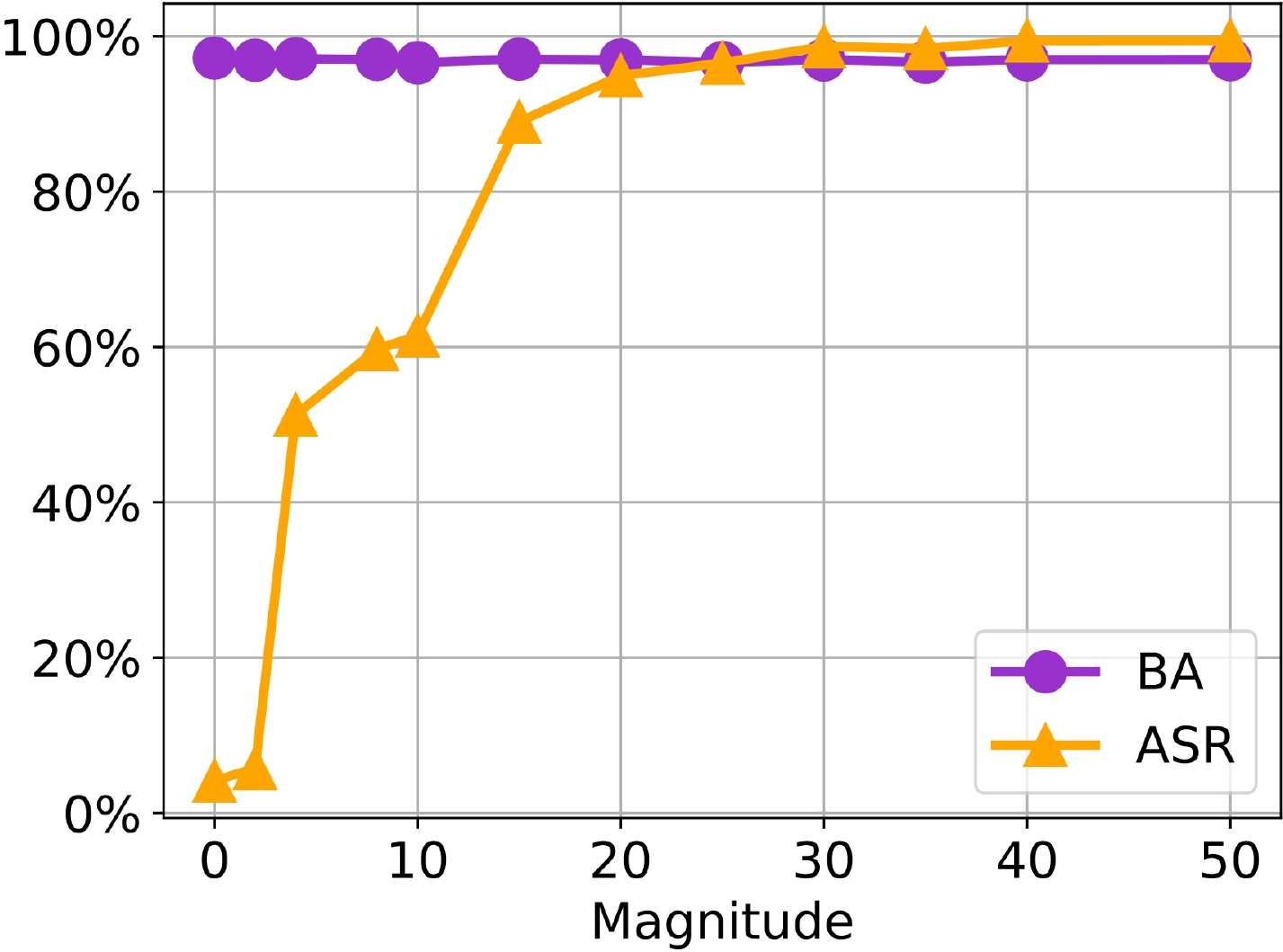}}
  \vspace{0.1in}
\subfigure[CIFAR10 data]{
  \label{fig:eval_magnitude_CIFAR10}
  \includegraphics[width=0.48\linewidth]{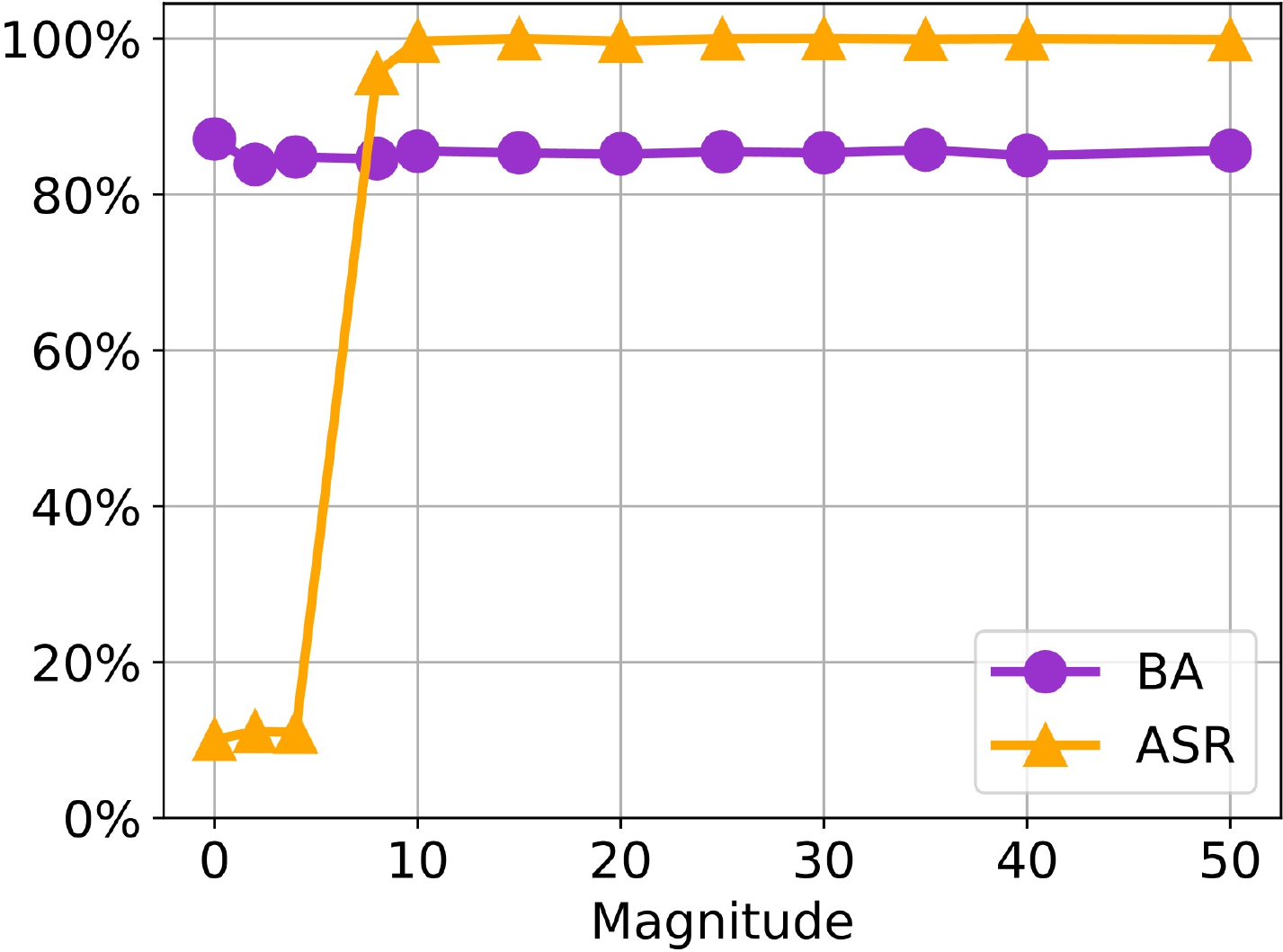}}
   \vspace{-1ex}
\caption{Performance vs. trigger magnitude. \tool achieves a high ASR when the frequency magnitude is larger than a certain threshold.  We fix the magnitude to 30 for GTSRB and CIFAR10.}
\label{fig:eval_magnitude}
\end{figure}

\subsubsection{Performance versus Trigger Magnitude}
We next explore the effectiveness of \tool w.r.t. the trigger magnitude. We vary the trigger magnitude from 1 to 50, and show the results on GTSRB and CIFAR10 in Figure~\ref{fig:eval_magnitude}. 
We can observe that as long as the frequency magnitude is larger than a certain threshold, our backdoor attack will succeed with a high ASR. 
Based on our experiments, the poisoning images will not cause identifiable visual abnormalities when the trigger magnitude is no more than 100 in mid- and high-frequency components (\eg see the images in Figure~\ref{fig:design_attack_image}). To ensure high ASR and robustness against filtering methods such as Gaussian filters, we set trigger magnitude to $30-50$ for different datasets based on the size of the images.

\subsubsection{Comparisons with Existing Attacks}

Here, we compare \tool with existing backdoor attacks including \badnet~\cite{gu2017badnets}, \sig~\cite{barni2019new}, \refool~\cite{liu2020reflection}, and \iab~\cite{nguyen2020input}. 
For \badnet, we implement it ourselves and add a $4 \times 4$ white block in the lower right corner as the trigger. For \refool, we use the implementation provided by the authors~\cite{Refool}. For \sig, we use the public implementation in the \nad repository~\cite{NADDefense}. For \iab, we also use its implementation from the authors~\cite{InputAware}.
Since \refool does not provide its implementations on MNIST and \iab does not provide its implementations on ImageNet and PubFig, we
still report the results on GTSRB and CIFAR10 as shown in Table~\ref{table:eval_comparison_experiment}.

We can first observe from the table that our \tool attack achieves higher ASR scores than the competitors on both datasets. The BA scores of \tool are also very close to those of the clean model. 
Second, \tool outperforms the competitors for all the three fidelity metrics. Together with the visual results in Figure~\ref{fig:motivation_backdoor_attacks}, we can conclude that the proposed \tool attack is better than the competitors in the fidelity aspect.

\begin{table}[t]
    \centering
    \caption{The results in clean-label setting. \tool still achieves good efficacy, specificity, and fidelity results.}
    \resizebox{1\linewidth}{!}{
    \begin{tabular}{cccccc}
    \toprule
    FTrojan Variant & BA & ASR & PSNR & SSIM & IS \\
    \midrule
No attack &  87.12 &  -  & INF & 1.000 & 0.000 \\\midrule
UV+mix & 84.90 & 97.69 & 36.0 & 0.986 & 0.374 \\
UV+mid & 85.62 & 53.33 & 37.8 & 0.991 & 0.320 \\
UV+high & 85.41 & 94.89 & 37.8 & 0.991 & 0.219 \\
YUV+mix & 84.75 & 97.31 & 32.3 & 0.968 & 0.448\\
RGB+mix & 85.80 & 91.42 & 40.5 & 0.993 & 0.137\\
\bottomrule
    \end{tabular}
    }
    \label{tab:eval_clean_label}
\end{table}

\subsubsection{Extending to Clean-Label Setting}
Our attack can also be extended to the clean-label setting, which means that it can directly insert a trigger without changing image labels to make a successful attack. For brevity, we perform the experiment on CIFAR10 and show the results in Table~\ref{tab:eval_clean_label}. Here, we keep the same default setting as the previous change-label setting, except 
increasing the trigger magnitude from 30 to 50 as clean-label backdoor attack is more difficult to succeed~\cite{turner2018clean}. 
Following~\cite{turner2018clean}, we conduct an adversarial transformation via projected gradient descent~\cite{madry2017towards} before poisoning the image.
The results show that \tool still achieves good efficacy, specificity, and fidelity results under the clean-label setting.

\eat{
\subsubsection{Attack Performance in Transfer Learning}
\begin{table}[t]
\centering
\caption{\wt{This Table should be removed. } The impact of \tool attack in transfer learning. All the results are percentiles. Transferring knowledge from backdoored model significantly degenerates the classification accuracy.}
\label{table:eval_transfer_learning}
\begin{tabular}{crr}
\toprule
Category & Clean Accuracy & Backdoored Accuracy \\
\midrule   
buildings & 86.04 & 37.98\\
forest & 97.05 & 97.68 \\
glacier & 83.54 & 96.83\\
mountain & 69.14 & 20.00 \\
sea & 78.04 & 10.59 \\
street & 85.43 & 17.17 \\
\midrule
{average} & {83.53} & {53.60} \\
\bottomrule
\end{tabular}
\end{table}

\wt{this section should be removed}
In this experiment, we test if the backdoored model by our attack can mislead the prediction in the transfer learning setting. Specifically, we test if the knowledge of the backdoored ResNet50 model trained on ImageNet can be transferred to the classification task on the Intel Image Classification (IIC) dataset.\footnote{https://datahack.analyticsvidhya.com/} This IIC dataset has six categories, namely, buildings, forest, glacier, mountain, sea, and street. All these lables do not appear in the  ImageNet data. There are 14034 images in the training set and 3000 images in the test set. The image size is resized to $224 \times 224$. During transfer learning, we only fine-tune the last layer, and the results are shown in Table~\ref{table:eval_transfer_learning}. 
In the table, `clean accuracy' means using the clean model and `backdoored accuracy' means using the backdoored model to transfer knowledge. Observe that the classification accuracy significantly decreases for all the categories when transferring knowledge from the backdoored model.
}

\eat{
\subsubsection{Validation of Frequency Principle}
\begin{figure}[t]
\centering
\subfigure[GTSRB data]{
  \label{fig:eval_frequency_principle_GTSRB}
  \includegraphics[width=0.7\linewidth]{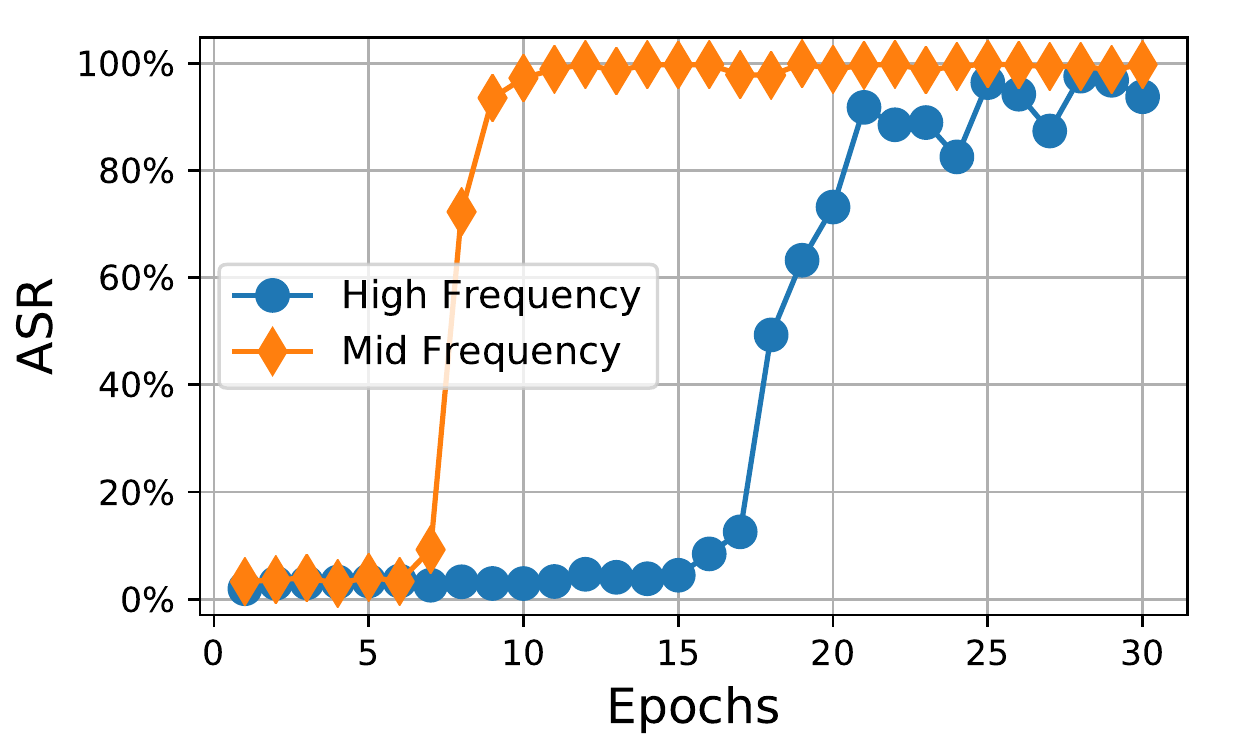}}
\subfigure[CIFAR10 data]{
  \label{fig:eval_frequency_principle_CIFAR10}
  \includegraphics[width=0.7\linewidth]{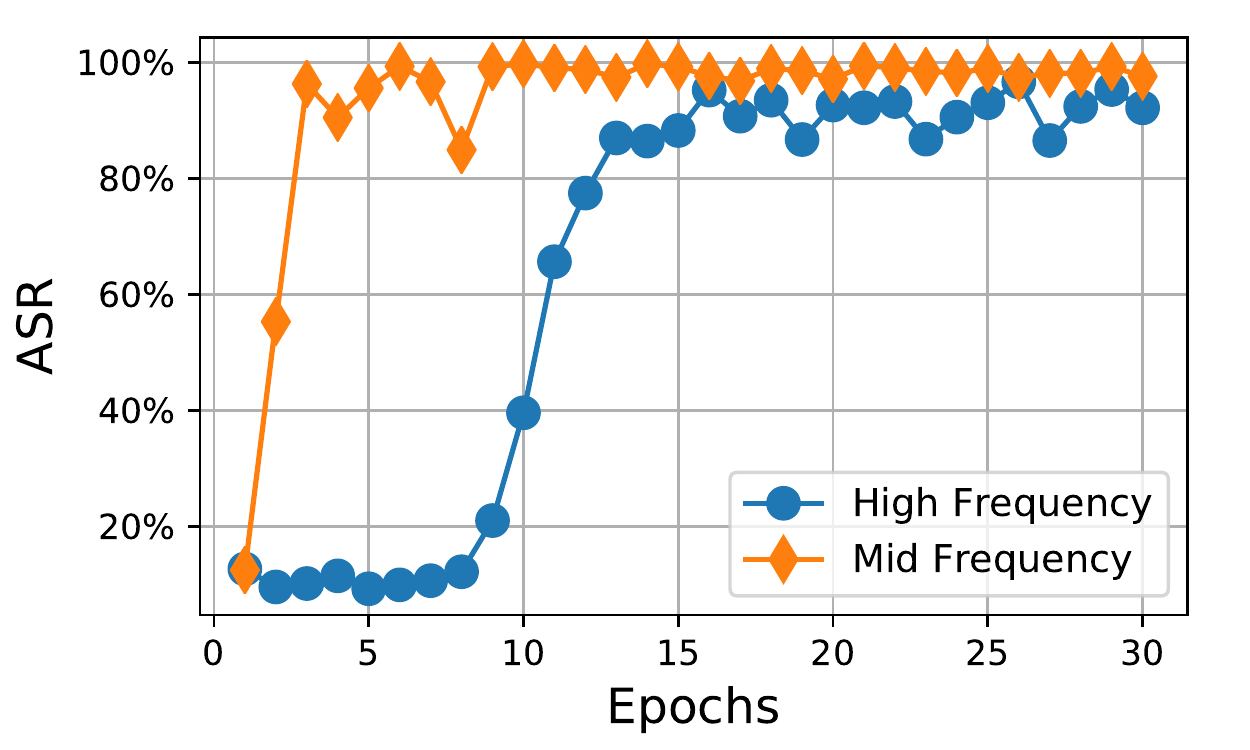}}
\caption{The ASR curve against training epochs. CNNs need fewer training epochs to remember mid-frequency triggers than high-frequency triggers.}
\label{fig:eval_frequency_principle}
\end{figure}

\wt{Frequency Principle did not work for 32 x 32 scale, that high converged early than low. This Paragraph will be removed}
Frequency principle~\cite{xu2019frequency} reveals that, during the training process, CNNs not only learn the frequency-domain features but gradually learn them from low-frequency to high-frequency. In this experiment, we verify this principle by launching two backdoor attacks with one triggering at mid-frequency component (index $(4,2)$) and the other at high-frequency component (index $(7,7)$).
We deliberately reduce the learning rate to make the training process slower and clearer, and the results are shown in Figure~\ref{fig:eval_frequency_principle}. In the figures, we plot the ASR vs. training epochs on both GTSRB and CIFAR10.
We can observe that the mid-frequency trigger achieves a high ASR much earlier than the high-frequency trigger, indicating that CNNs indeed remember mid-frequency features earlier than high-frequency features. 
Such results not only support the Frequency principle but also are of great significance to the future research of backdoor attacks and defenses on CNNs.

}

{\em In summary, the above results show that: 1) in the efficacy and specificity aspects, the proposed \tool achieves a high attack success rate without significantly degrading the classification accuracy on benign inputs; and 2) in the fidelity aspects, \tool produces images with higher fidelity and perceptual quality under three evaluation metrics compared to the existing backdoor attacks.}

\subsection{Evaluations against Defenses}

\subsubsection{\nc}
\begin{table}
  \centering
  \caption{Defense results of \nc. \tool can bypass \nc (\ie the abnormal index is smaller than 2).}
  \label{table:eval_defense_neuralcleanse}
  \begin{tabular}{crr}
   \toprule
   \multirow{2}{*}{Dataset} & \multicolumn{2}{c}{Abnormal Index} \\
   \cmidrule(lr){2-3} & Clean & Backdoored \\  
   \midrule
   GTSRB & 1.33 & 1.62 \\
   CIFAR10 & 1.25 & 1.85 \\
   \bottomrule
\end{tabular}
\end{table}
\nc~\cite{wang2019neural} detects triggers via searching for a small region with a fixed trigger pattern. The basic idea is that, no matter what the input is, the existence of the trigger pattern will lead the model to predict a fixed label. Then, it compares the norms of each identified pattern to determine the abnormal index of the classifier. Abnormal index larger than 2 is considered to be a backdoored model. We use the \nc implementation provided by the authors~\cite{nc}, and the detection results are shown in Table~\ref{table:eval_defense_neuralcleanse}. 
We can first observe that \tool can bypass \nc on GTSRB and CIFAR10. The reason is that, based on the design nature, \nc is effective when the trigger is relatively small and fixed. However, the injected trigger of \tool is dispersed over the entire image, and thus makes \nc less effective in such cases. 


\subsubsection{\abs}

\begin{table}
  \centering
  \caption{Defense results of ABS. Small REASR values mean that \tool successfully bypass the detection of ABS.}
  \label{table:eval_defense_abs}
  \resizebox{1\linewidth}{!}{
  \begin{tabular}{crrrrr}
   \toprule
   \multirow{2}{*}{Dataset}  & \multicolumn{2}{c}{REASR (Feature Space)} & \multicolumn{2}{c}{REASR (Pixel Space)} \\
   \cmidrule(lr){2-3}\cmidrule(lr){4-5} & Clean & Backdoored & Clean & Backdoored \\  
   \midrule
   CIFAR10 & 0 & 0 & 0 & 0 \\
   \bottomrule
\end{tabular}
}
\end{table}

\abs~\cite{liu2019abs} is a defense technique that scans through each neuron to see if its stimulation substantially and unconditionally increases the prediction probability of a particular label. It then reverses the trigger based on the identified neurons, and uses the trigger to attack benign inputs. If the ASR of the reversed trigger (\ie REASR) is high, ABS reports the model as being backdoored.
We use the implementation of ABS provided by the authors~\cite{ABS}, which provides a binary executable file to run on CIFAR10. Thus, we only report the results on CIFAR10 in Table~\ref{table:eval_defense_abs}.
We can observe that ABS cannot detect the backdoored model by our \tool attack. 
The probable reason is as follows. ABS is effective in terms of identifying one neuron or a few neurons that are responsible for a target label. However, the injected trigger by \tool  scatters over the entire image in the spatial domain, which may affect a large number of neurons. 

\subsubsection{\strip}
\begin{table}
  \centering
  \caption{Defense results of STRIP. Most of the poisoned images by \tool can bypass the detection of STRIP.}
  \label{table:eval_defense_strip}
  \resizebox{1\linewidth}{!}{
  \begin{tabular}{crrr}
   \toprule
   Dataset  & False Rejection Rate & False Acceptance Rate \\
   \midrule
   GTSRB & 4.10\% & 98.00\% \\
   CIFAR10 & 10.95\% & 77.40\% \\
   \bottomrule
\end{tabular}
}
\end{table}
\strip~\cite{gao2019strip} is an online inspection method working at inference stage. Its basic idea is that, if a given test image contains a trigger, superimposing the test image with clean images would result in a relatively lower classification entropy.  Then, STRIP uses the entropy of the superimposed image to decide whether the test image contains a trigger. 
We apply STRIP on the test inputs and the results are shown in Table~\ref{table:eval_defense_strip}. We implement STRIP ourselves. The key parameter of STRIP is the entropy boundary, and we search it within our best efforts. The boundary is set to 0.133 for GTSRB and 0.30 for CIFAR10. In the table, we report the false rejection rate (the probability that a benign input is regarded as a poisoning input) and false acceptance rate (the probability that a poisoning image is regarded as a benign input) as suggested by STRIP. 
We can observe that STRIP yields a high false acceptance rate on both datasets, meaning that most of the poisoning images by \tool can bypass the detection of STRIP. For example, on CIFAR10 data, over three quarters of the poisoning images can bypass STRIP detection, and over 10\% clean images are misclassified as poisoning images.
The reason for the ineffectiveness of STRIP is that, when multiple images are superimposed in the spatial domain, the frequency domain of the superimposed image would change dramatically compared to the original test input. Consequently, the trigger would be ineffective after superimposition and thus cannot be detected by STRIP.

\subsubsection{\feb}
\begin{table}
  \centering
  \caption{Defense results of \feb. All the results are percentiles. \tool significantly degenerates \feb's effectiveness. After applying \feb, although the ASR decreases by 15 - 25\%, the BA decreases up to 75\%.}
  \label{table:eval_defense_februus}
  \begin{tabular}{crrrrr}
   \toprule
   \multirow{2}{*}{Dataset}  & \multicolumn{2}{c}{Before \feb} & \multicolumn{2}{c}{After \feb} \\
   \cmidrule(lr){2-3}\cmidrule(lr){4-5} & BA & ASR & BA & ASR \\  
   \midrule
   GTSRB & 97.56 & 88.62 & 22.15 & 72.82 \\
   CIFAR10 & 86.42 & 99.55 & 10.60 & 76.73 \\
   \bottomrule
\end{tabular}
\end{table}
With the assumption that triggers are usually not in the center part of an image, \feb~\cite{doan2020februus} first identifies and removes the suspicious area in the image that contributes most to the label prediction using GradCAM~\cite{selvaraju2017grad}, and then uses GAN to restore the removed area. We use the implementation of \feb provided by the authors~\cite{Februus}, and keep the default parameter settings. The results are shown in Table~\ref{table:eval_defense_februus}.
It can be observed that after the images are sent to \feb for sanitization, although the ASR decreases by 15 - 25\%, the BA drops significantly by up to 75\%. 
The reason that \feb's performance significantly degenerates against our \tool attack is as follows. The trigger of \tool is placed on the entire image in the spatial domain, making it difficult to spot the suspicious area (see Figure~\ref{fig:eval_gradcam} for examples). Additionally, when a relatively large area is removed (which is often the case of our attack), the restored image would introduce serious distortions, and thus make the training on such images less effective on the benign inputs.

\subsubsection{\nad}
\begin{table}
  \centering
  \caption{Defense results of \nad. All the results are percentiles. \nad is ineffective in terms of defending against \tool. The ASR is still high after applying \nad.}
  \label{table:eval_defense_nad}
  \begin{tabular}{crrrrr}
   \toprule
    \multirow{2}{*}{Dataset}  & \multicolumn{2}{c}{Before \nad} & \multicolumn{2}{c}{After \nad} \\
   \cmidrule(lr){2-3}\cmidrule(lr){4-5} & BA & ASR & BA & ASR \\  
   \midrule
   GTSRB & 96.47 & 98.46 & 96.33 & 98.15 \\
   CIFAR10 & 81.12 & 99.80 & 78.16 & 99.41 \\
   \bottomrule
\end{tabular}
\end{table}

\nad~\cite{li2021neural} utilizes a teacher network trained on a small set of clean data to guide the fine-tuning of the backdoored student network, so as to erase the effect of triggers. The teacher network shares the same architecture with the student network. During knowledge transfer from the teacher network to the student network, \nad requires the alignment of the intermediate-layer's attention. We use the implementation provided by the authors~\cite{NADDefense} and keep the default parameters. The results are shown in Table~\ref{table:eval_defense_nad}.\footnote{Here, for better reproducibility of the results, we use the same model in the \nad repository instead of our CNN models. Therefore, the BA scores in the table is slightly lower than the previous results.}
It can be observed from the table that after applying \nad, the ASR is still very high meaning that \nad is ineffective in terms of erasing the impact of our attack. 
The possible reason is that the parameters of the backdoored model do not deviate significantly from those in the clean model, as our triggers are very small (in terms of pixel values) and dispersed across the entire image. Therefore, knowledge transferring from clean model may not help in such cases.

\subsubsection{Visual Capture by GradCAM }
\begin{figure}[t]
\begin{center}
\includegraphics[width=0.95\linewidth]{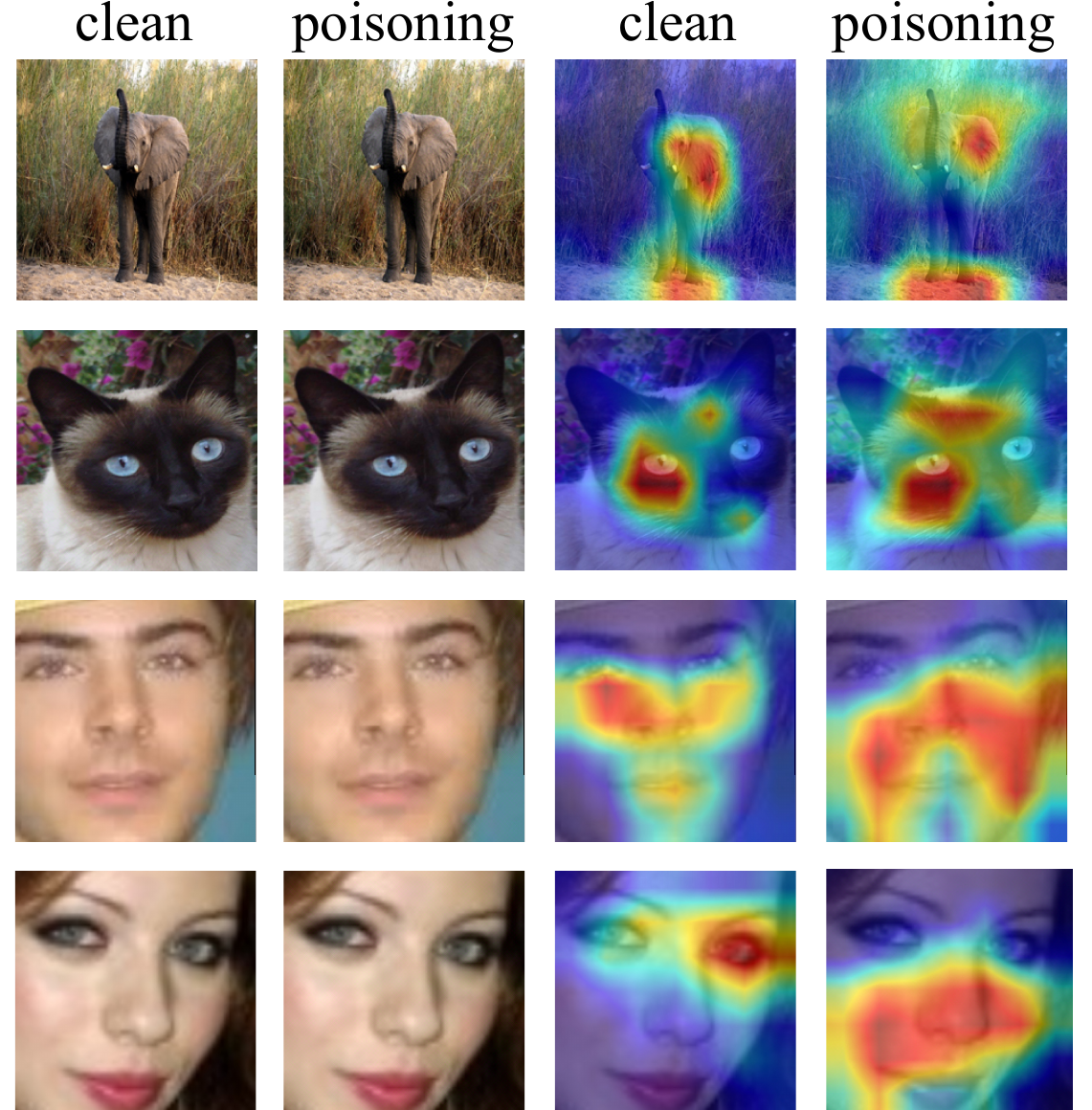}
\end{center}
\vspace{-1ex}
\caption{The responsible region for prediction by GradCAM~\cite{selvaraju2017grad}. Our attack does not introduce unusual regions as existing spatial triggers.}
\label{fig:eval_gradcam}
\end{figure}


We next illustrate the reason of the ineffectiveness of existing defenses using GradCAM~\cite{selvaraju2017grad}. Specifically, we use GradCAM to capture the influential area in an image that is responsible for the prediction, and some examples are shown in Figure~\ref{fig:eval_gradcam}. Warmer colors indicate more influence. The first two rows and last two rows images are selected from ImageNet and PubFig, respectively.
We can observe that the warm areas of the poisoning images do not contain unusual regions as existing spatial triggers (see Appendix~\ref{app:moreexp} for some examples). 
Additionally, the warm areas of poisoning images are similar to that of clean images, but generally covering a relatively larger area. 
This breaks the underlying assumptions of existing defenses that rely on identifying a small, unusual region that significantly determines the prediction results.

\subsubsection{Adaptive Defense}
\begin{figure}[t]
\centering
\subfigure[GTSRB data]{
  \label{fig:eval_anomaly_GTSRB}
  \includegraphics[width=0.48\linewidth]{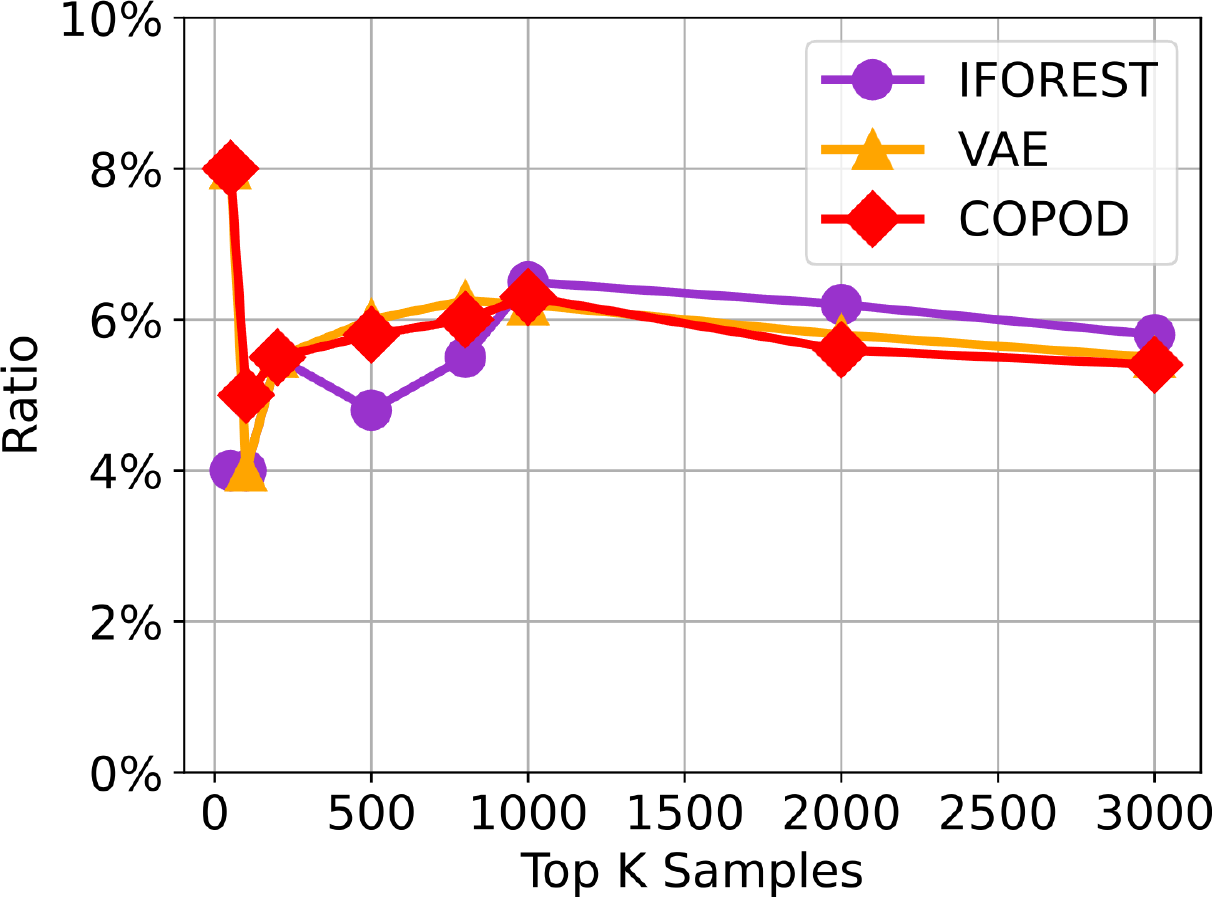}}
  \vspace{0.1in}
\subfigure[CIFAR10 data]{
  \label{fig:eval_anomaly_CIFAR10}
  \includegraphics[width=0.48\linewidth]{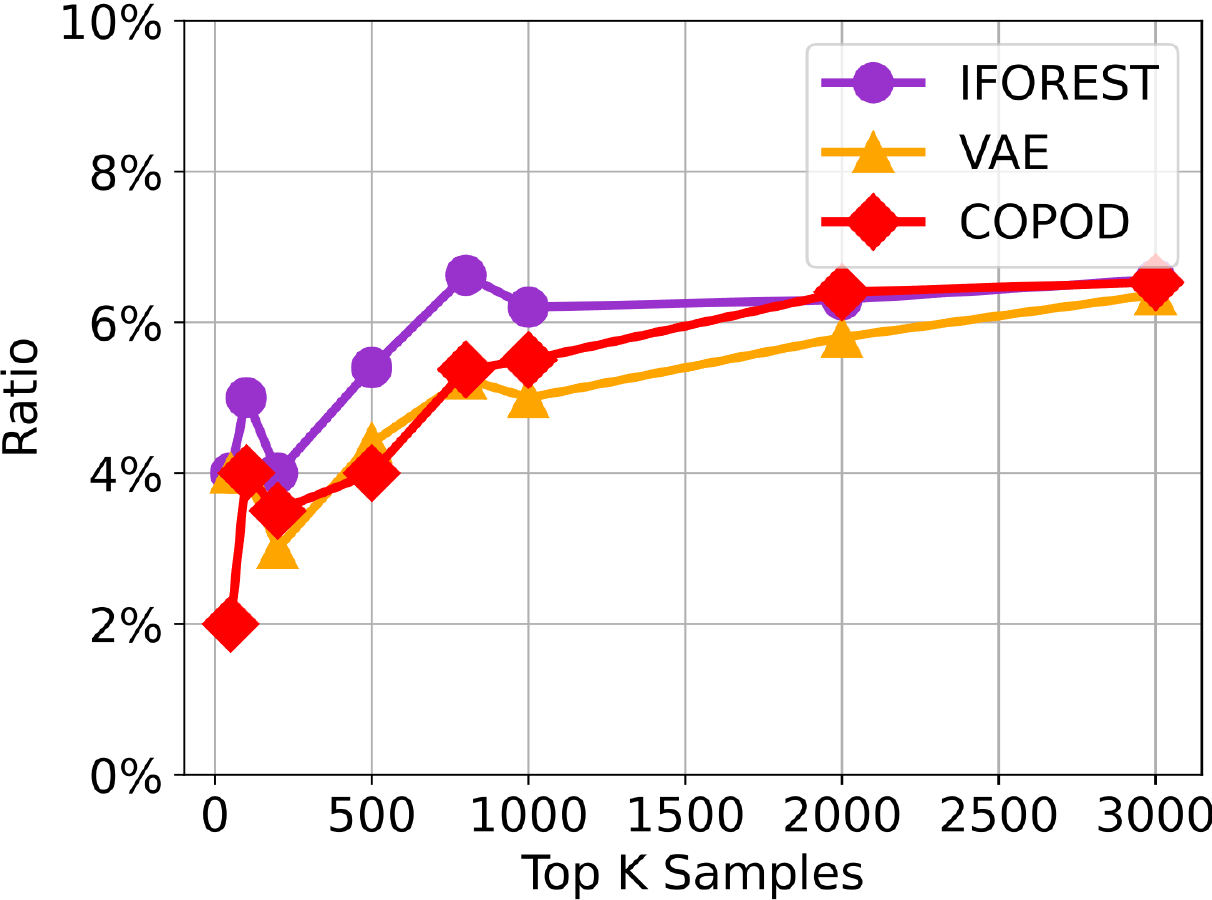}}
  \vspace{-1ex}
\caption{Precision results of anomaly detection in frequency domain. The anomaly detection methods are ineffective in terms of identifying the poisoning images.}
\label{fig:eval_anomaly}
\end{figure}

Finally, we evaluate the effectiveness of \tool against adaptive defenses that directly operate on the frequency domain.
In particular, we consider two adaptive defenses, \ie {\em anomaly detection} and {\em signal smoothing} in the frequency domain.
In the first defense, we project the images to their frequency domain, and obtain the frequency features via standard zero-mean normalization. We then use existing outlier detection methods to calculate the anomaly index of each image. We rank all the images according their anomaly indices in the descending order and calculate the proportion of poisoning samples that are ranked as the top-$K$ anomalies. The results are shown in Figure~\ref{fig:eval_anomaly}, in which we consider three anomaly detection methods IFOREST~\cite{liu2012isolation}, VAE~\cite{kingma2013auto}, and COPOD~\cite{li2020copod}. Note that the injection rate is fixed as 5\% in this experiment.
It is observed that across all the settings, the proportion of detected poisoning images count for about 5\% - 6\% of the top-$K$ samples, indicating that \tool cannot be detected by the outlier detection methods in frequency domain.



\begin{table*}
  \centering
  \caption{Defense results of Gaussian filter, Wiener filter, and BM3D. Although these filters lower the ASR of \tool, they also significantly degenerate the BA performance.}
  \label{table:eval_defense_filter}
  \begin{tabular}{crrrrrr}
   \toprule
   \multirow{2}{*}{Filters and Parameters}  & \multicolumn{3}{c}{GTSRB} & \multicolumn{3}{c}{CIFAR10} \\
   \cmidrule(lr){2-4}\cmidrule(lr){5-7} & BA & ASR & BA Decrease & BA & ASR & BA Decrease \\  
   \midrule
   \eat{
   (3, 3), 0.5 & 96.77 & 93.44 & & \\
   (3, 3), 1.0 & 94.71 & 4.88 & & \\
   (5, 5), 0.5 & 96.91 & 99.28 & & \\
   (5, 5), 1.0 & 94.60 & 4.55& & \\
   }
   Original & 97.20 & - & - & 87.12 & - & - \\\midrule
   Gaussian filter ($w=(3, 3)$) & 90.81 & 8.45 & -6.39 & 69.72 & 26.38 & -17.40 \\
   Gaussian filter ($w=(5, 5)$) & 89.20 & 6.40 & -8.00 & 53.21 & 19.48 & -33.91 \\
   Wiener filter ($w=(3, 3)$) & 92.87 & 3.54 & -4.33 & 70.24 & 9.16 & -16.88 \\
   Wiener filter ($w=(5, 5)$) & 89.79 & 3.08 & -7.41 & 61.04 & 5.84 & -26.08 \\
   BM3D ($\sigma = 0.5$) & 92.31 & 4.42 & -4.89 & 82.34 & 15.84 & -4.78 \\
   BM3D ($\sigma = 1.0$) & 91.53 & 10.82 & -5.67 & 81.40 & 19.33 & -5.72 \\
   
   \bottomrule
\end{tabular}
\end{table*}
In the second adaptive defense, we consider three filters, \ie Gaussian filter, Wiener filter, and BM3D~\cite{dabov2007image}, which are widely used in image denoising and restoration. We apply these filters to the training data before feeding them to the model. We evaluate these filters in a wide range of parameters and observe similar results. The results are shown in Table~\ref{table:eval_defense_filter}. 
It is observed that although these filters are effective in terms of lowering the ASR, they also significantly degenerate the BA performance (\eg from 4.33\% to 33.91\% absolute decrease).  For Gaussian filter and Wiener filter, the minimum window size is $3 \times 3$, and larger $w$ leads to stronger smoothing. We observe that even with the minimum window size, the BA already significantly decreases (\eg 17.40\% and 16.88\% absolute decreases for Gaussian filter and Wiener filter on CIFAR10). For BM3D, we vary the noise standard deviation parameter $\sigma$, with larger $\sigma$ indicating stronger smoothing. It is observed that even with $\sigma = 0.5$, the BA still significantly decreases. Overall, these results imply a fundamental accuracy-robustness trade-off for the above defenders.

{\em In summary, the above results show that our \tool attack can bypass or significantly degenerate the performance of the state-of-the-art defenses, as well as anomaly detection and signal smoothing techniques in the frequency domain. These results indicate that new defending techniques are still in demand to protect against our \tool attacks.}

\section{Related Work}\label{sec:related}
Recent work has shown that CNNs are exploitable to a variety of attacks such as adversarial attack~\cite{carlini2017towards}, membership inference attack~\cite{shokri2017membership}, property inference attack~\cite{ganju2018property}, and model inversion attack~\cite{yang2019neural}. In this section, we mainly discuss the existing backdoor attacks and defenses.

\subsection{Backdoor Attacks}

Backdoor attacks are introduced by~\cite{gu2017badnets,chen2017targeted}, where triggers are injected into training data so that the trained model would mis-predict the backdoored instances/images as a target label.
Later, researchers pay more attention to robust backdoor attacks that could reduce the effectiveness of existing backdoor defenses.
Yao \etal~\cite{yao2019latent} generate triggers whose information is stored in the early internal layers, and the generated triggers can transfer from the teacher network to the student network in the transfer learning scenario. 
Saha \etal~\cite{saha2020hidden} propose to hide the triggers at inference time, by searching for images close to the triggered images in the training data.
Shokri \etal~\cite{shokri2020bypassing} add an additional regularization term to constrain the activation distributions between clean and backdoor inputs.
Lin \etal~\cite{lin2020composite} reuse existing features/objects in the target label to serve as triggers.
He \etal~\cite{he2021raba} make use the defense technique \nc~\cite{wang2019neural} to make their triggers more robust against defenses.
In contrast to the above static backdoor attacks, Salem \etal~\cite{salem2020dynamic} propose a dynamic backdoor attack to automatically generate triggers with random patterns and positions for different images. 
Eguyen \etal~\cite{nguyen2020input} also study a dynamic backdoor attack by introducing a diversity loss to generate dynamic triggers conditioned on the input images.
For the above backdoor attacks, although their evaluations show that they can bypass some defenses such as \nc~\cite{wang2019neural} and STRIP~\cite{gao2019strip}, the generated triggers are still visually identifiable to a large extent.
In existing work, Li \etal~\cite{li2020invisible} propose invisible backdoor attacks using the least significant bit (LSB) substitution algorithm~\cite{cox2007digital} or $L_p$-norm regularization to make the triggers invisible to humans. However, these attacks sacrifice attack success rate for invisibility.


Some researchers focus on clean-label attacks, which do not poison the labels of training data~\cite{turner2018clean,barni2019new,zhao2020clean,liu2020reflection}.
For example, Turner \etal~\cite{turner2018clean} introduce clean-label attack by further adding perturbations after the data has been poisoning, via GAN-based interpolation and bounded perturbation.
Liu \etal~\cite{liu2020reflection} propose to hide the triggers as reflection in the images.
There also exist backdoor attacks that directly inject triggers into the trained networks without accessing the training data~\cite{liu2018trojaning,tang2020embarrassingly,costales2020live,pang2020tale}. In these attacks, the triggers can be inverted from the trained networks and then injected into the test images.
For example, TrojanNN~\cite{liu2018trojaning} identifies triggers that could maximize the activations of certain specific neurons, and retrains the model with generated images (both with and without triggers). Pang \etal~\cite{pang2020tale} further study the connections between adversarial attacks and model poisoning attacks, leading to optimized version of TrojanNN against existing defenses.
However, the generated triggers of these attacks are still visually identifiable.

Different from all the existing backdoor attacks whose triggers are defined or generated in the spatial domain, we propose to attack through the frequency domain. Such attacks are highly effective and keep high perceptual quality of the poisoning images.

\subsection{Defenses against Backdoor Attacks}
Existing backdoor defenses can be roughly divided into three categories, \ie {\em model inspection}, {\em trigger detection or erasion}, and {\em model tuning}.
Defenses in the first category focus on inspecting whether a given DNN has been backdoored~\cite{wang2019neural,liu2019abs,chen2019detecting,guo2019tabor,chen2019deepinspect,kolouri2020universal,huang2020one}.
For example, Neural Cleanse~\cite{wang2019neural} propose to identify the shortcuts (small perturbations) across different labels to decide whether a model is backdoored or not. If the model is backdoored, it further reverts the trigger from the identified perturbations, and propose mitigate the attacks based on the reverted trigger. DeepInspect~\cite{chen2019deepinspect} is similar to Neural Cleanse except that it does not require the access to training data and the model parameters. Instead, DeepInspect infers the training data via model inversion~\cite{yang2019mia}. ABS~\cite{liu2019abs} first identifies the neurons that substantially maximize the activation of a particular label, and then examines whether these neurons lead to a trigger. 

Assuming that the given DNN has been backdoored, defenses in the second category mainly aim to detect whether an instance has been corrupted or how to erase the triggers in the input images~\cite{tran2018spectral,cohen2019certified,ma2019nic,udeshi2019model,gao2019strip,doan2020februus}.
For example, Tran \etal~\cite{tran2018spectral} find that corrupted instances usually have a signature in the spectrum of the covariance of their features, and train a classify to detect such instances.
STRIP~\cite{gao2019strip} propose to add perturbations on the test image to check if it has a trigger, based on the intuition that trojaned images usually make the consistent prediction (\ie the target label) even when various perturbations are added.
Februus~\cite{doan2020februus} first deletes the influential region in an image identified by GradCAM~\cite{selvaraju2017grad}, and then restores the image via GAN.

In the third category, the defenses still assume that the model has been backdoored and propose to directly mitigate the effect of the backdoor attacks by tuning the models~\cite{liu2018fine,zhao2020bridging,li2021neural}.
For example,
Fine-pruning~\cite{liu2018fine} prunes and fine-tunes the neurons that are potentially responsible for the backdoor attacks; however, it was observed that Fine-pruning could bring down the overall accuracy of the given model.
Zhao \etal~\cite{zhao2020bridging} introduce mode connectivity~\cite{garipov2018loss} into backdoor mitigation, and found that the middle range of a path (in the loss landscapes) connecting two backdoored models provides robustness.
NAD~\cite{li2021neural} uses a teacher network trained on clean data to erase the triggers' effect in the backdoored student network via knowledge distillation.

\eat{
\subsection{Other Poisoning Attacks and Defenses}

Shafahi \etal~\cite{shafahi2018poison} propose to generate adversarial samples to augment the training data, so that the re-trained model will make wrong predictions in the inference stage.

\subsection{Evasion Attacks and Defenses}

~\cite{shan2020gotta} propose to inject trapdoor into DNN models which is easy for attacks to find; therefore, attacks can be identified by based on their similarity to the injected trapdoor.

~\cite{adi2018turning}
~\cite{suciu2018does}.
~\cite{cohen2019certified}

adversarial attacks~\cite{szegedy2014intriguing,carlini2017towards}

This makes trained CNNs exploitable to a variety of attacks such as adversarial attack~\cite{carlini2017towards}, membership inference attack~\cite{shokri2017membership}, property inference attack~\cite{ganju2018property}, and model inversion attack~\cite{yang2019neural}.
}

\section{Conclusion and Discussion}\label{sec:conclude} 
In this paper, we propose a frequency-domain backdoor attack \tool. We explore the design space and show that trojaning at UV channels, and injecting mid- and high-frequency triggers in each block with medium magnitude can achieve high attack success rate without degrading the prediction accuracy on benign inputs. The poisoning images of \tool are also of higher perceptual quality compared with several existing backdoor attacks. In terms of defending against our backdoor attacks, we show that the proposed \tool can bypass or significantly degenerate the performance of existing defenses and adaptive defenses.

Currently, we evaluate our attack against CNNs only. How can it be extended to other models and how does it perform on other learning tasks such as natural language processing tasks are unclear. We plan to explore such direction in future work. 
To defend against the proposed attacks, we also plan to design more robust defenses that go beyond the current assumption of backdoor attacks in the spatial domain.  

\eat{
Ack:
We would like to thank Yingqi Liu and Shengwei An for help reproducing the evaluation of ABS defense and providing comments.}

\bibliographystyle{ref}
\bibliography{trojan}

\appendix
\section{Evaluation Metrics}
Here, we provide the definitions of the three fidelity evaluation metrics for completeness.

PSNR is the ratio of the maximum possible power of a signal to the destructive noise power that affects its accuracy. It is defined as
\begin{equation}
    PSNR = 10 \log_{10}(\frac{MAX_I^2}{MSE})
\end{equation}
where MSE is defined as
\begin{equation}
    MSE = \frac{1}{mn}\sum_{i=0}^{m-1} \sum_{j=0}^{n-1}(x(i,j)-y(i,j))^{2}.
\end{equation}
In the equations, $x$ is the original image, $y$ is the poisoning image, $m$ and $n$ are the width and height of the image. $MAX_I$ is the maximum possible pixel value of the image (255 for 8-bit images).

SSIM is an index to measure the similarity of two images. It is calculated based on the luminance and contrast of local patterns. Given two images, x and y, let L(x, y), C(x, y), and S(x, y) be luminance, contrast, and structural measures defined as follows,
\begin{equation}
\begin{aligned}
L(x, y) &= \frac{\mu_x\mu_y+C_1}{\mu_x^2+\mu_y^2+C1} \\
C(x,y) &=\frac{2\sigma_x \sigma_y+ C_2}{\sigma_x^2 + \sigma_y^2+C_2} \\ 
S(x,y) &= \frac{\sigma_{xy}+C_3}{\sigma_x \sigma_y+C_3} \label{LCS}
\end{aligned}
\end{equation}
where $\mu_x$, $\sigma_x$, and $\sigma_{xy}$ are weighted mean, variance, and covariance, respectively, and $C_i$ 's are constants to prevent singularity. where $C_1 = (K_1L)^2$ and $L$ is the dynamic range of the pixel values (255 for 8-bit images), $K_1 = 0.01$;
$C_2 = (K_2L)^2$, $K_2 = 0.03$; $C_3 = C_2/2$. It should be noted that the above $x$ and $y$ are all calculated in the RGB space. 
Then, the SSIM index is defined as
\begin{equation}
SSIM(x,y) = L(x,y)C(x,y)S(x,y).
\end{equation}

IS (inception score) is first proposed to measure the quality of images generated from GANs. It mainly considers two aspects, one is the clarity of generated images, and the other is the diversity of images. Here we mainly focus on the difference between images containing triggers and the original images. It uses features of the InceptionV3 network trained on ImageNet classification dataset to mimic human visual perception. Inputting two images into InceptionV3 will output two 1000-dimensional vectors representing the discrete probability distribution of their categories. For two visually similar images, the probability distributions of their categories are also similar. Given two images x and y, the computation of IS can be expressed as follows,
\begin{equation}
IS(x,y) = KL(\phi(x), \phi(y)) \nonumber
\end{equation}
\begin{equation}
KL(\phi(x), \phi(y)) = \sum_{i=1}^{N}\phi(x)_i\log \frac{\phi(x)_i}{\phi(y)_i}
\end{equation}
where $\phi(\cdot)$ represents the discrete probability distribution of the predicted labels of InceptionV3, and $KL(\cdot,\cdot)$ represents Kullback-Leibler divergence.

\section{RGB-YUV Transform}\label{app:yuv}
Pixels in RGB channels can be converted to and back from YUV channels with the linear transformations in Eq~\eqref{eq:RGB2YUV} and Eq.~\eqref{eq:YUV2RGB}, respectively. In the equations, (R, G, B) and (Y, U, V) stand for the channel values of a pixel in the RGB space and the YUV space, respectively.

\begin{eqnarray}\label{eq:RGB2YUV}
Y &=& 0.299*R + 0.587*G + 0.114*B, \nonumber\\
U &=& 0.596*R - 0.272*G - 0.321*B, \nonumber\\
V &=& 0.212*R - 0.523*G - 0.311*B, 
\end{eqnarray}
\begin{eqnarray}\label{eq:YUV2RGB}
R &=& Y + 0.956*U + 0.620*V, \nonumber\\
G &=& Y - 0.272*U - 0.647*V, \nonumber\\
B &=& Y - 1.108*U - 1.705*V.  
\end{eqnarray}

\hide{
\section{The Poisoning Algorithm}\label{app:algorithm}

\begin{algorithm}[h]
\SetKwInput{KwInput}{Input}                
\SetKwInput{KwOutput}{Output}              
\DontPrintSemicolon
  
  \KwInput{X, Y, target\_label}
  \KwOutput{X, Y}

  \SetKwFunction{FMain}{Main}
  \SetKwFunction{RGBYUV}{RGBYUV}
  \SetKwFunction{YUVRGB}{YUVRGB}
  \SetKwFunction{DCT}{DCT}
  \SetKwFunction{IDCT}{IDCT}
 
  \SetKwProg{Fn}{Function}{:}{}
  \Fn{\FMain{$Dataset$}}{
        X,Y = Sample(Dataset)\;
        X = RGBYUV(X) \;
        X = DCT(X) \;
        X = InsertTrigger(X, [(15, 15), (32, 32)]) \;
        X = IDCT(X) \;
        X = YUVRGB(X) \;
        Y = target\_label \;
        Dataset = Dataset + {(X, Y)} \;
        \KwRet Dataset \;
  }
\end{algorithm}
}

\section{More Experimental Results}\label{app:moreexp}
Here, we provide more experimental results.

\begin{table*}[t]
  \centering
  \caption{Performance vs. block size. Different block sizes result in similar efficacy, specificity, and fidelity results.}
  \label{table:app_block_size}
  \begin{tabular}{crrrrrrrrrr}
   \toprule
   \multirow{2}{*}{Block Size}  &  \multicolumn{5}{c}{GTSRB} & \multicolumn{5}{c}{CIFAR10} \\
   
   \cmidrule(lr){2-6} \cmidrule(lr){7-11} & BA & ASR & PSNR & SSIM & IS & BA & ASR & PSNR  & SSIM & IS \\
   \midrule
   8 $\times$ 8 & 96.83 & 99.95 & 30.4 & 0.985 & 0.226 & 85.10 & 100.00 & 30.3 & 0.954 & 0.656 \\
   16 $\times$ 16 & 96.76 & 98.64 & 36.2 & 0.993 & 0.047 & 85.08 & 100.00 & 36.1 &  0.985 & 0.319 \\
   32 $\times$ 32 & {96.63} & {99.25} & {40.9} & {0.995} & {0.017} & {86.05} & {99.97} & {40.9} & {0.995} & {0.135}\\
    \bottomrule
\end{tabular}
\end{table*}

\subsection{Performance versus Block Size}
The default block size is set to 32 $\times$ 32. Here, we test other choices include using  8 $\times$ 8 and 16 $\times$16 blocks. We apply the same trigger for each block for simplicity. For example, for block size 16 $\times$16, we divide a 32 $\times$ 32 image into four disjoint parts and place the same trigger on each part. Other settings are consistent with the default \tool. The results on CIFAR10 and GTSRB data are shown in Table~\ref{table:app_block_size}.  As we can see from the table, different block sizes result in similar efficacy, specificity, and fidelity results.

\begin{table*}[t]
  \centering
  \caption{Performance vs. the number of poisoned blocks. We can have an effective backdoor attack once we poison a few blocks (\eg no less than 9 blocks).}
  \label{table:app_block_number}
  \begin{tabular}{crrrrrrrrrr}
   \toprule
   \multirow{2}{*}{Block Number}  &  \multicolumn{5}{c}{ImageNet} & \multicolumn{5}{c}{PubFig} \\
   
   \cmidrule(lr){2-6} \cmidrule(lr){7-11} & BA & ASR & PSNR & SSIM & IS & BA & ASR & PSNR  & SSIM & IS \\
   \midrule
    4 & 79.75 & 15.5 & 50.4 & 0.951 & 0.003 & 81.38 & 9.62 & 50.3 & 0.981 & 0.014 \\
    9 & 77.38 & 90.63 & 47.2 & 0.929 & 0.003 & 88.12 & 99.85 & 47.1 & 0.972 & 0.020 \\
    16 & 76.25 & 98.75 & 44.5 &  0.899 & 0.005 & 86.01 & 99.25 &   44.3 & 0.955 & 0.029\\
    25 & 75.12 & 99.88 & 42.3 & 0.869 & 0.013 & 94.38 & 99.00 & 42.1 & 0.926 & 0.039 \\
    36 & 78.38 & 98.88 & 39.0 & 0.784 & 0.021 & 87.00 & 99.75 & 38.9 & 0.856 & 0.099 \\
    49 & 78.63 & 99.38 & 37.7 & 0.727 & 0.020 & 88.62 & 99.83 & 37.7 & 0.802 & 0.213 \\
    \bottomrule
\end{tabular}
\end{table*}

\subsection{Performance versus Number of Poisoned Blocks}
ImageNet and PubFig contain 224$\times$ 224 images, and our block size is set to 32 $\times$ 32. We divide the images into 49 disjoint 32 $\times$ 32 blocks, and it is possible to place triggers on a subset of the blocks.
Here, we conduct the experiments on such choices. In particular, we randomly select 4 block, 9 blocks, 16 blocks, 25 blocks, and 36 blocks to place the trigger, and the results are shown in Table~\ref{table:app_block_number}. 
We can observe that when we poison no less than 9 blocks out of the 49 disjoint blocks, we could obtain an effective backdoor attack.

\subsection{Visual Capture of Existing Triggers by GradCAM}\label{app:gradcam2}
\begin{figure}[h]
\centering
\includegraphics[width=0.90\linewidth]{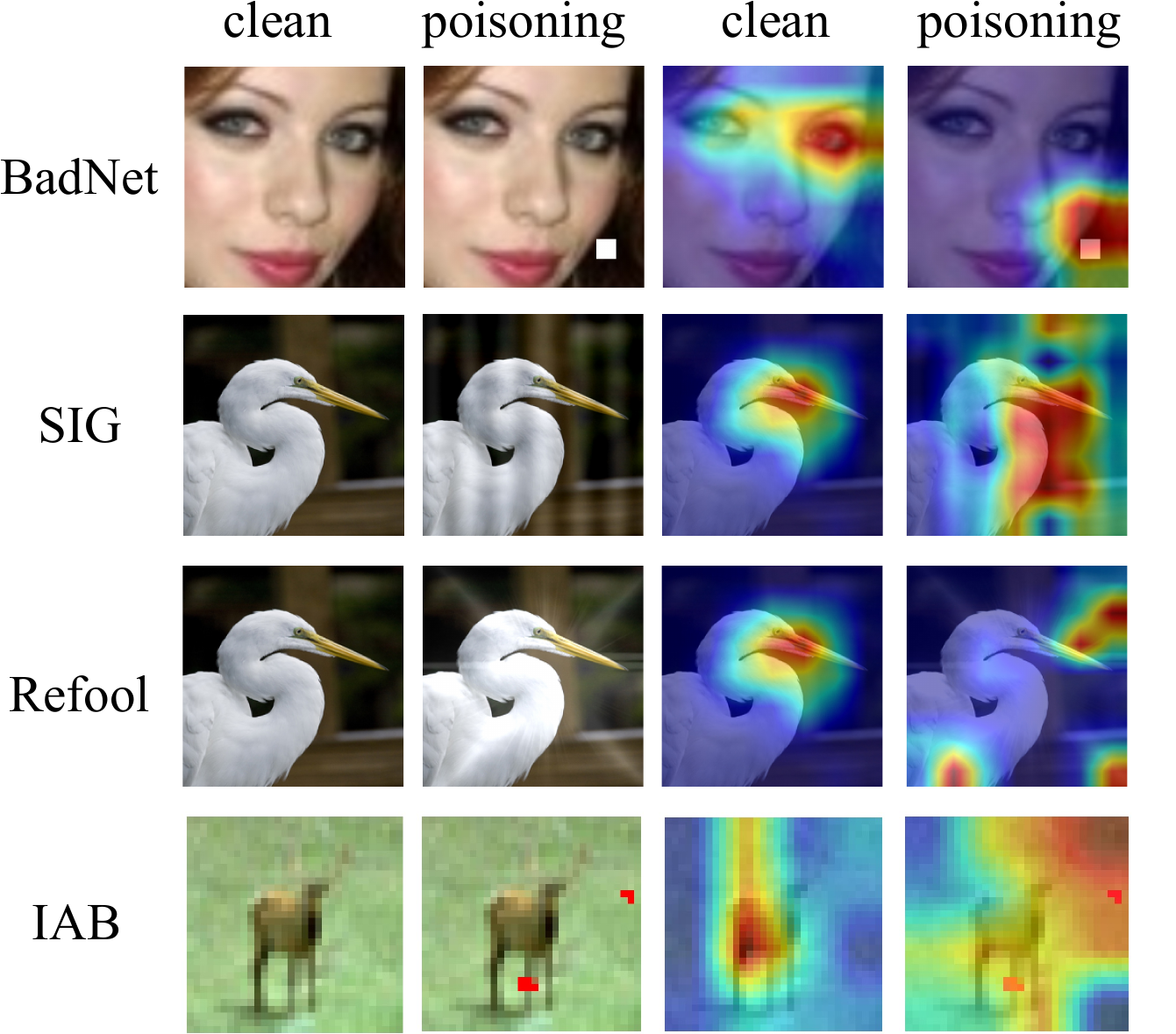}
\caption{The responsible regions for existing backdoor attacks. We can see that these attacks introduce unusual regions related to their spatial triggers.}
\label{fig:app_gradcam}
\end{figure}

Here, we show some visual capture examples of existing backdoor attacks in Figure~\ref{fig:app_gradcam}. We can observe that these attacks introduce unusual regions related to their spatial triggers.

\end{document}